\documentclass{caps}
\usepackage{epsfig}

\def\msun{{\rm M_{\odot}}}
\def\rsun{{\rm R_{\odot}}}

      \DeclareSymbolFont{UPM}{U}{eur}{m}{n}
      \SetSymbolFont{UPM}{bold}{U}{eur}{b}{n}
      \DeclareSymbolFont{AMSa}{U}{msa}{m}{n}
      \DeclareMathSymbol{\upi}{0}{UPM}{"19}
      \DeclareMathSymbol{\umu}{0}{UPM}{"16}
      \DeclareMathSymbol{\upartial}{0}{UPM}{"40}
      \DeclareMathSymbol{\leqslant}{3}{AMSa}{"36}
      \DeclareMathSymbol{\geqslant}{3}{AMSa}{"3E}
      \DeclareMathSymbol{\la}{3}{AMSa}{46}
      \DeclareMathSymbol{\ga}{3}{AMSa}{38}

      \let\leq=\leqslant

\begin{document}

\pagenumbering{roman}
\tableofcontents
\cleardoublepage

\pagenumbering{arabic}

\author[A.R. King]{A.R. KING\\Theoretical Astrophysics Group,
University of Leicester, Leicester LE1 7RH, UK}
\chapter{Accretion in Compact Binaries}

\section{Introduction}

Compact binaries have long been a paradigm for accretion theory. Much
of our present view of how accretion occurs comes directly from the
comparison of theory with observations of these sources. Since theory
differs little for other objects such as active galaxies, increasing
efforts have recently gone into searching for correspondences in
observed behaviour. This chapter aims at giving a concise summary of
the field, with particular emphasis on new developments since the
previous edition of this book. 

These developments have been significant. Much of the earlier
literature implicitly assumed that accreting binaries were fairly
steady sources accreting most of the mass entering their vicinity,
often with main--sequence companions, and radiating the resulting
accretion luminosity in rough isotropy. We shall see that in reality
these assumptions fail for the majority of systems. Most are
transient; mass ejection in winds and jets is extremely common; a
large (sometimes dominant) fraction of even short--period systems have
evolved companions whose structure deviates significantly from the
zero--age main sequence; and the radiation pattern of many objects is
significantly anisotropic. It is now possible to give a complete
characterization of the observed incidence of transient and persistent
sources in terms of the disc instability model and formation
constraints. X--ray populations in external galaxies, particularly the
ultraluminous sources, are revealing important new insights into
accretion processes and compact binary evolution.

\section{Accretion disc theory}

Essentially all of the systems discussed here accrete via discs.
Accretion disc theory is the subject of many books and reviews (see
e.g. Frank et al., 2002 and Pringle, 1981). Accordingly this section
simply summarizes the main results without giving detailed
derivations.

\subsection{Disc formation}

Matter accreting on to a mass $M_1$ forms a disc if its specific
angular momentum $J$ is too large for it to impact the object
directly. We define the circularization radius
\begin{equation}
R_{\rm circ} = {J^2\over GM_1},
\label{1}
\end{equation}
which is where the matter would orbit if it lost energy but no angular
momentum. The condition for disc formation is typically that $R_{\rm
circ}$ should exceed the effective size of the accretor (a parabolic
orbit with specific angular momentum $J$ would reach a minimum
separation $0.5R_{\rm circ}$). This effective size is identical with
the radius of a non--magnetic white dwarf or neutron star, but is of
order the magnetospheric radius if there is a dynamically significant
magnetic field. For a black hole the effective size is the radius of
the last stable circular orbit. If mass transfer occurs via Roche lobe
overflow, $J$ is comparable with the specific orbital angular momentum
of the binary, $R_{\rm circ}$ is large, and the condition for disc
formation is almost always satisfied in a compact binary. The
exceptions are some cataclysmic variables where the white dwarf
accretor is strongly magnetic, and double white dwarf systems with
companion/accretor mass ratios $M_2/M_1$ larger than about 0.15 (see
e.g. Nelemans et al., 2001).

In the usual case that matter can indeed orbit at $R_{\rm circ}$, disc
formation will follow under the assumption that energy is lost through
dissipation faster than angular momentum is redistributed. Since the
orbit of lowest energy for a given angular momentum is a circle,
matter will follow a sequence of circular orbits about the compact
accretor. The agency for both energy and angular momentum loss is
called viscosity. For many years the nature of this process was
mysterious, but recently a strong candidate has emerged, in the form
of the magnetorotational instability (MRI: Balbus \& Hawley,
1991). Here a comparatively weak magnetic field threading the disc is
wound up by the shear, and transports angular momentum
outwards. Reconnection limits the field growth and produces
dissipation. Numerical simulations show that this is a highly
promising mechanism, and will shortly reach the point of allowing
direct comparison with observations.

\subsection{Thin discs}

While viscosity transports angular momentum and thus spreads the
initial ring at $R_{\rm circ}$ into a disc, the nature of this
accretion disc is determined by the efficiency with which the disc can
cool. In many cases this is high enough that the disc is {\it thin}:
that is, its scaleheight $H$ obeys
\begin{equation}
H \simeq {c_{\rm s}\over v_{\rm K}}R << R
\label{2}
\end{equation}
at disc radius $R$, where $c_{\rm s}$ is the local sound speed, and 
\begin{equation} 
v_{\rm K} = \biggl({GM\over R}\biggr)^{1/2}
\label{3}
\end{equation}
is the Kepler velocity, with $M$ the accretor mass. In this state the
azimuthal velocity is close to $v_{\rm K}$, and the radial and
vertical velocities are much smaller. The properties of being thin,
Keplerian and efficiently cooled are all equivalent, and if any one of
them breaks down so do the other two.

If the thin disc approximation holds, the vertical structure is almost
hydrostatic and decouples from the horizontal structure, which can be
described in terms of its surface density $\Sigma$. If the disc
is axisymmetric, mass and angular momentum conservation imply that the
latter obeys a nonlinear diffusion equation
\begin{equation}
{\partial\Sigma\over\partial t} = {3\over R}{\partial\over\partial R} 
\biggl(R^{1/2}{\partial\over\partial R}[\nu\Sigma R^{1/2}]\biggr).
\label{4}
\end{equation}
Here $\nu$ is the kinematic viscosity, which is usually parametrized
as
\begin{equation}
\nu = \alpha c_{\rm s}H.
\label{5}
\end{equation}
where $\alpha$ is a dimensionless number. In a steady state this gives
\begin{equation}
\nu\Sigma = {\dot M\over 3\pi}\biggl[1 - \beta\biggl({R_{\rm in}\over
R}\biggr)^{1/2}\biggr], 
\label{6}
\end{equation}
where $\dot M$ is the accretion rate and the dimensionless quantity
$\beta$ is specified by the boundary condition at the inner edge
$R_{\rm in}$ of the disc. For example, a disc ending at the radius
$R_*$ of a non--rotating star has $R_{\rm in} = R_*$. In a steady thin
disc dissipation $D(R)$ per unit surface area is also proportional to
$\nu\Sigma$, i.e.
\begin{equation}
D(R) = {9\over 8}\nu\Sigma{GM\over R^3}\biggl[1 - \beta\biggl({R_{\rm in}\over
R}\biggr)^{1/2}\biggr], 
\label{7}
\end{equation}
so that the surface temperature $T$ is independent of the viscosity
$\nu$ despite being entirely generated by it:
\begin{equation}
T = T_{\rm visc} = \left\{\frac{3GM\dot{M}}{8\pi R^{3}\sigma}
\left[1-\beta\left(\frac{R_{\ast}}{R}\right)^{1/2}\right]\right\}^{1/4}.
\label{8}
\end{equation}

\subsection{Disc timescales}

Equation (\ref{4}) shows that $\Sigma$ changes
on a timescale
\begin{equation}
t_{\rm visc} \sim {l^2\over \nu}
\label{9}
\end{equation}
if its spatial gradient is over a lengthscale $l$. Hence we would
expect a disc to make significant changes in its surface density and
thus its luminosity on a timescale $\sim R^2/\nu$, where $R$ is its
outer radius. We can use this fact to get an idea of the magnitude of
the viscosity in observed discs. In dwarf novae, which are
short--period white--dwarf binaries, the disc size is $R \sim 1 -
3\times 10^{10}$~cm, and surface density changes take a few days. This
suggests that $\alpha \sim 0.1$. Encouragingly, numerical simulations
of the MRI give comparable answers. There are two other obvious
timescales in a disc. The first is the dynamical timescale
\begin{equation}
t_{\rm dyn} \sim {R\over v_{\rm K}} = \biggl({R^3\over
GM}\biggr)^{1/2},
\label{10}
\end{equation}
characterizing states in which dynamical equilibrium is disturbed;
note that vertical hydrostatic balance is resored on a timescale
\begin{equation}
t_z \sim {H\over c_{\rm s}} = {R\over v_{\rm K}} = t_{\rm dyn}
\label{11}
\end{equation}
where we have used eqn (\ref{2}). The second is the thermal timescale
\begin{equation}
t_{\rm th} = {\Sigma c_{\rm s}^2\over D(R)} \sim {R^3c_{\rm s}^2\over
GM\nu} = {c_{\rm s}^2\over v_{\rm K}^2}{R^2\over \nu} = \biggl({H\over
R}\biggr)^2 t_{\rm visc}
\label{12}
\end{equation}
where we have used eqn (\ref{7}). The alpha--disc parametrization
(\ref{4})
can be used to show that
\begin{equation}
t_{\rm visc} \sim {1\over \alpha}\biggl({H\over R}\biggr)^{-2}t_{\rm dyn}
\label{13}
\end{equation}
so we finally have the ordering
\begin{equation}
t_{\rm dyn} \sim t_{z} \sim \alpha t_{\rm th} \sim \alpha(H/R)^{2} t_{\rm visc},
\label{14}
\end{equation}
i.e. dynamical $<$ thermal $<$ viscous.

\subsection{Breakdown of the thin disc approximation}
\label{ad}

The thin disc approximation discussed above requires the accreting
matter to cool efficiently. However flows with low radiative
efficiency on to a black hole can in principle occur, for at least two
reasons: the accretion rate $\dot M$ may be so low that the inflowing
gas has low density and thus a long cooling time, or conversely $\dot
M$ may be so large that the flow is very optically thick, and
radiation is trapped and dragged down the hole. As energy is advected
inwards, these flows are called ADAFs (advection--dominated accretion
flows). If the accretor is not a black hole, the advected energy must
be released near the surface of the accretor. This effect has been
invoked to explain observations of quiescent transients (see Section
\ref{hor} below). 

Considerable theoretical effort has gone into trying to understand
such radiatively inefficient flows. Clearly the flows cannot be
geometrically thin, making analytic treatments difficult. Such studies
assume a discontinuous change from a thin Keplerian disc to an ADAF at
some `transition radius' $R_{\rm tr}$, whereas it is easy to show that
any transition region must be extended itself over a size several
times the assumed $R_{\rm tr}$. Moreover the difficulty of applying a
predictive theory for the disc viscosity means that $R_{\rm tr}$ is
taken as a free parameter in attempts to fit observations, a freedom
that would not be present in reality. Recent numerical studies of
accretion at low radiative efficiency (e.g. Stone, Pringle \&
Begelman, 1999, Stone \& Pringle, 2001) find that very little of the
matter flowing in at large radius actually accretes to the black
hole. Instead the time--averaged mass inflow and outflow rates both
increase strong with radius, and almost cancel. The simulations have
not so far been run for long enough to reach any kind of steady state:
one obvious possibility is that the density may eventually reach
values at which radiative cooling does become efficient, leading to a
thin disc phase.

\subsection{Warping of discs}

An important effect in disc physics, discovered only recently
(Pringle, 1996) is that accretion discs tend to warp if exposed to
irradiation from a source at their centres, which may be the accretion
flow itself. The origin of the warping is that the disc must scatter
or reradiate the incident radiation, which results in a pressure force
normal to its surface. If the surface is perturbed from complete
axisymmetry the force can increase or decrease in such a way as to
cause the perturbation to grow. As the gravitational potential is
close to spherical symmetry near the accreting object, it is quite
possible for disc material to orbit at angles to the binary plane. A
full perturbation analysis (Pringle 1996) shows that the condition for
warping is
\begin{equation}
L \ga 12\pi^2\nu_2\Sigma v_{\phi}c,
\label{warp}
\end{equation}
where $\nu_2$ is the vertical kinematic viscosity coefficient.

This inequality shows that we can expect warping in discs in
sufficiently luminous systems. We can re--express (\ref{warp}) 
by defining the ratio of the vertical viscosity
coefficient $\nu_2$ to the usual radial one $\nu$ as $\psi =
\nu_2/\nu$. Then (\ref{6}) allows us to write (\ref{warp}) as
\begin{equation}
L \ga 12\pi^2\psi\nu\Sigma v_{\phi}c = 4\pi\psi\dot Mv_{\phi}c
\label{warp2}
\end{equation}
for a steady disc. Now
\begin{equation}
v_{\phi} = \biggl({GM\over R}\biggr)^{1/2} = \biggl({R_{\rm Schw}\over
2R}\biggr)^{1/2}c,
\end{equation}
where $R_{\rm Schw} = 2GM/c^2$ is the Schwarzschild radius of the
central star,
so combining with (\ref{warp2}) gives the condition
\begin{equation}
L \ga 4\pi\psi\dot M c^2\biggl({R_{\rm Schw}\over 2R}\biggr)^{1/2}.
\label{warp3}
\end{equation}
If we finally assume that the central luminosity $L$ comes entirely
from accretion at the steady rate $\dot M$ on to a compact object of
radius $R_*$, we can write
\begin{equation}
L \simeq \dot M c^2 {R_{\rm Schw}\over R_*},
\end{equation}
and use (\ref{warp3}) to give
\begin{equation}
{R\over R_*} \ga 8\pi^2\psi^2{R_*\over R_{\rm Schw}}.
\label{warp4}
\end{equation}

Equation (\ref{warp4}) now tells us if warping is likely in various
systems. First, it is clearly very unlikely in accretion--powered
white dwarf binaries such as CVs, since with $M= 1\msun, R_* = 5\times
10^8$~cm and $\psi \sim 1$ we find the requirement $R \ga 7\times
10^{13}$~cm, demanding binary periods of several years. (Warping may
occur in some supersoft X--ray binaries, where the energy source is
nuclear burning rather than the gravitational energy release of the
accreted matter.) For neutron stars and black holes by contrast
warping is probable if the disc is steady, since with $M = 1\msun, R_*
= 10^6$~cm, $\psi \sim 1$ we find $R \ga 3\times 10^8$~cm, while for a
black hole with $R_* = R_{\rm Schw}$ warping will occur for $R \ga
8\pi^2R_{\rm Schw} \sim 2.4\times 10^7(M/1\msun)$~cm. It thus seems
very likely that discs in LMXBs are unstable to warping, at least for
persistent systems. A warped disc shape therefore offers an
explanation for the observed facts that persistent LMXB discs are both
irradiated and apparently have large vertical extent. The X--ray light
curves of persistent LMXBs are strongly structured, implying that the
X--rays are scattered by the accretion flow White \&
Holt (1982).

Of course, the discussion above only tells us about the possible onset
of warping. To see what shape the disc ultimately adopts one must
resort to numerical calculations. These have so far only been
performed in a highly simplified manner, but do provide suggestive
results. For example, one might imagine that warps would be
self--limiting, in that a significant warp would geometrically block
the very radiation driving the warp. Numerical calculations show that
this does not occur: the reason is that warping always starts at large
disc radii (cf \ref{warp4}), where the mean disc plane is perturbed
away from the original one. Matter flowing inwards from these radii
has angular momentum aligned with the perturbed disc plane, and so
transfers a `memory' of it to the inner disc. One can show that a
warped disc always has a line of nodes following a leading spiral,
with the result that the disc can become markedly distorted from its
original plane shape without shadowing large areas and
arresting the growth of the warp. Of course these calculations are
highly simplified, and one might in reality expect the warps to be
limited (perhaps by tidal torques, before attaining such
distorted shapes. However it is clear that radiation--induced warping
is a promising mechanism for making persistent LMXB discs deviate from
the standard picture. An attractive feature of this explanation is
that, although the discs deviate {\it globally} from the standard
picture, the thin disc approximation nevertheless continues to apply
{\it locally}, in the sense that the warping is always over
lengthscales much larger than the local scaleheight $H$.

\subsection{Accretion disc stability}

Many accreting sources are observed to vary strongly. The clearest
examples are dwarf novae (DN), which are binaries in which a white dwarf
accretes from a low mass star, and soft X--ray transients (SXTs),
where a black hole or neutron star accretes from a low--mass
companion. In both cases the system spends most of its time in
quiescence, with occasional outbursts in which it is much
brighter. However, beyond this qualitative similarity, there are very
clear quantitative differences. In dwarf novae the typical timescales
are: quiescence $\sim$ weeks -- months, outburst $\sim$ days, and the
system luminosity typically rises from $\sim 10^{32}$~erg~s$^{-1}$ to
$\sim 10^{34}$~erg~s$^{-1}$. In SXTs, the corresponding numbers are
quiescence $\sim$ 1 -- 50 yr or more, outburst $\sim$ months, system
luminosity rises from $\sim 10^{32}$~erg~s$^{-1}$ to 
$10^{38} - 10^{39}$~erg~s$^{-1}$. Remarkably, it is possible to explain both
types of system with a similar model: the SXT version contains only
one extra ingredient over that currently accepted for DN.

The basic model at work in both cases is the disc instability
picture. There is a huge literature on this subject: Lasota (2001) and
Frank et al. (2002) give recent reviews. The fundamental idea behind
the model is that in a certain range of mass transfer rates, the disc
can exist in either of two states: a hot, high viscosity state
(outburst) and a cool, low viscosity state (quiescence). In practice
these two states correspond to hydrogen existing in ionized or neutral
states respectively. The very steep dependence of opacity on
ionization fraction and thus temperature makes any intermediate states
unstable, and the disc jumps between the hot and cool states on a
thermal timescale. In each of these two states it evolves on a viscous
timescale. The heirarchy (\ref{14}) shows that this pattern does
qualitatively reproduce the observed behaviour of long quiescence,
short outburst, with rapid transitions between them. Since the basic
cause of instability is hydrogen ionization we can immediately deduce
the condition for a disc to be stable: it must have no ionization
zones. Thus a sufficient condition for suppressing outbursts and
making a system persistent is that its surface temperature $T$ should
exceed some value $T_{\rm H}$ characteristic of hydrogen ionization (a
typical value for $T_{\rm H}$ is 6500 K, depending somewhat on the disc
radius). We can assume that this condition is also necessary if the
system is to be persistent, i.e. a system is persistent if and only if
\begin{equation}
T_{\rm visc} > T_{\rm H} 
\label{15}
\end{equation}
throughout its accretion disc. Since $T$ decreases with disc radius
(cf eqn \ref{8}) this condition is most stringent at the outer disc
edge $R = R_{\rm out}$, so we require
\begin{equation}
T_{\rm visc}(R_{\rm out}) > T_{\rm H} 
\label{16}
\end{equation}
for stability. In principle there is another family of
persistent sources where the opposite condition
\begin{equation}
T_{\rm visc} < T_{\rm H} 
\label{17}
\end{equation}
holds throughout the disc; the condition is tightest close to the
inner edge of the disc, where $T$ has a maximum value. There is
clearly a strong selection effect against finding such systems, which
must be inherently faint, and do not call attention to themselves by
having outbursts. 

In the following sections we shall see that the occurrence of
outbursts implies powerful constraints on the evolution of both CVs
and X--ray binaries.

\section{Dwarf novae: the nature of the outbursts}

The disc instability picture described above works quite well when
applied to dwarf novae (see the review by Lasota, 2001), given the
limitations of current treatments of disc viscosity. There is good
reason to hope that improvements here will refine the picture
further. In particular, two-- and three--dimensional disc simulations
have given a realistic picture of many disc phenomena which remained
obscure in the early one--dimensional calculations. The first example
of this was Whitehurst's (1988) investigation of superhumps.
Superhumps are a photometric modulation at a period slightly longer
than the spectroscopically determined orbital period. They occur in a
subclass of short--period dwarf novae during particularly long
outbursts called superoutbursts. They had defied many attempted
explanations until Whitehurst's simulations of discs residing in the
full Roche potential, rather than simply the field of the accreting
star. These revealed that in binaries with sufficiently small
secondary--to--primary mass ratios $q = M_2/M_1 \la 0.25$ the disc becomes
eccentric and precesses progradely within the binary. Tidal stressing
of this disc causes dissipation and thus the superhump modulation.

This picture explains many of the observed superhump
properties. Because the modulation results from intrinsic tidal
stressing rather than geometrical effects, it is independent of the
system inclination. Given that white dwarfs in CVs have masses $M_1$
lying in a small range $\sim 0.6\msun - 1.0\msun$ the restriction to
small mass ratios $q$ means that $M_2$ must be small ($\la 0.15\msun -
0.25\msun$), accounting for the fact that almost all of the
superoutbursting systems have periods below the 2--3~hr CV gap. The
reason for the restriction to small $q$ was subsequently traced to the
fact that the superhump phenomenon is driven by the 3:1 orbital
resonance (Whitehurst \& King, 1991; Lubow, 1991); only for small
ratios can the disc get large enough to access this resonance.

For some time there were competing explanations for the superoutbursts
themselves. One (Vogt, 1983) invoked enhanced mass transfer from the
secondary triggered by a normal thermal--viscous instability. In
contrast Osaki (1989) suggested that in a series of normal outbursts
the disc grows in size because accretion on to the white dwarf
successively removes matter of low angular momentum. After several
such episodes the disc reaches the 3:1 resonant radius, where tides
remove angular momentum very effectively and cause more prolonged
accretion of a significant fraction of the disc mass. Recent 2--D
simulations (Truss et al., 2001) show that the latter model does
function as suggested, and agrees with observation. Superoutbursts
are a direct result of tidal instability. No enhanced mass transfer
from the secondary is required to initiate or sustain either the
superoutburst or the superhumps, provided that the mass ratio is small
enough that the disc can grow to the 3:1 resonant radius.

\section{Dwarf novae: the occurrence of the outbursts}

\subsection{Short--period dwarf novae}

An important question for the disc instability idea is whether it
correctly divides observed CV systems into dwarf novae and persistent
(novalike) systems. In particular all non--magnetic CVs with periods
below the well--known gap at 2--3 hr are dwarf novae. To answer this
question we have to predict conditions such as mass transfer rate
$-\dot M_2$ as a function of binary period $P$, and then check
condition (\ref{15}) by setting $\dot M = -\dot M_2$ in
(\ref{8}). Evidently the inequality (\ref{15}) should fail for systems
below the gap. In the standard view of CV evolution the secondary
stars are assumed to be completely unevolved low--mass main sequence
stars, and CV binaries evolve under angular momentum loss via
gravitational radiation and magnetic stellar wind braking (see
e.g. King, 1988 for a review). This picture leads to mass transfer
rates $-\dot M_2(P)$ which fulfil our expectation above (see
Fig. \ref{cvfig}): the disc instability picture correctly predicts
that short--period non--magnetic CVs are dwarf novae.

However things are clearly more difficult for systems above the
gap. Here there is a mixture of dwarf novae and novalikes, strongly
suggesting that the simple recipe described above for checking
(\ref{15}) does not capture the essence of the situation. As the disc
instability picture seems to describe outbursts quite well, given a
suitable mass transfer rate $-\dot M_2$, the most likely resolution of
the problem is that the adopted relation for $-\dot M_2(P)$ is too
simple. This problem is not confined to dwarf novae, but appears to be
generic to all CVs above the period gap.

One possibility uses the fact that $-\dot M_2(P)$ is the {\it
average} mass transfer rate, taken over timescales
$>10^5$~yr. Fluctuations on timescales shorter than this could still
be unobservable in any individual system, but lead to an effective
spread in instantaneous mass transfer rates. This suggestion has the
virtue of leaving intact the existing picture of long--term CV
evolution (e.g. the period histogram), which uses $\-\dot M_2(P)$. The
fluctuations might themselves result from cycles driven by irradiation
of the companion star (King et al., 1995). The main problem for this
approach is that it tends to predict an almost bimodal distribution of
instantaneous mass transfer rates, with the low state too low to give
dwarf nova properties in good agreement with observation.

A second quite different idea (King \& Schenker, 2002; Schenker \&
King, 2002) uses the observed fact that many CV secondaries have
spectral types significantly later than would be expected for a ZAMS
star filling the Roche lobe (Baraffe \& Kolb, 2000). The idea here is
that the spread in $-\dot M_2$ reflects real differences in the nature
of the secondary star; in many cases this has descended from a star
which was originally more massive than the white dwarf, allowing
significant nuclear evolution (I shall use the term `evolved' to
describe any star whose internal structure deviates significantly from
the zero--age main sequence, even if the exterior appearance resembles
a ZAMS star.)  Moreover the large mass ratio means that the donor
star's Roche lobe tends to contract as it loses mass, leading to mass
transfer on its thermal timescale.
\begin{figure}
  \begin{center}
    \epsfig{file=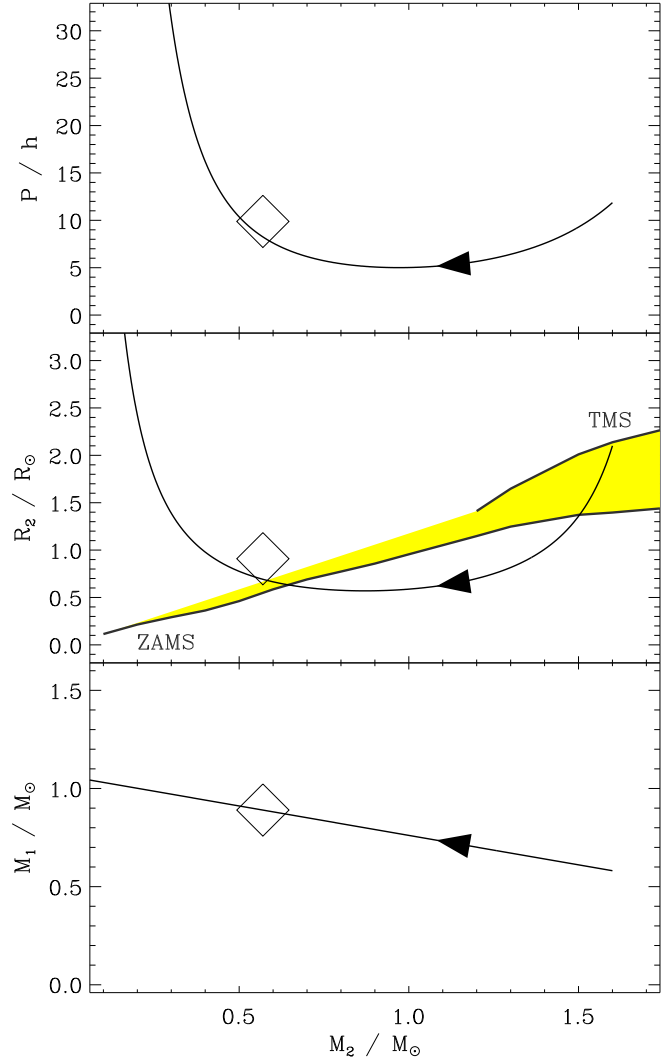, width=10cm}
  \end{center}
\caption{Evolution of a model AE Aqr progenitor system. For the
assumed parameters and $M_{1, \rm now} = 0.89\msun$ various tracks are
shown, starting from a $1.6\msun$ star which has almost reached its
maximum MS radius. The three panels show the evolution of the
orbital period, secondary radius and WD mass for different cases
of mass loss $\dot M = -\eta\dot M_2$ (dotted line: $\eta = 1$ --
full line: $\eta = 0$ -- dashed line: $\eta = 0.3$). The diamond marks
the current position of AE Aqr. Note that the conservative model
($\eta = 1$) cannot meet the requirements for $M_1$: even starting
from the lowest possible WD mass, it has grown well beyond the
Chandrasekhar limit before reaching the current $P_{\rm orb}$ of
AE~Aqr. The shaded area in the middle panel marks the radii of single
stars during their main--sequence life as labelled. (From Schenker et
al., 2002).}
\label{aefig}
\end{figure}
\begin{figure}
  \begin{center}
    \epsfig{file=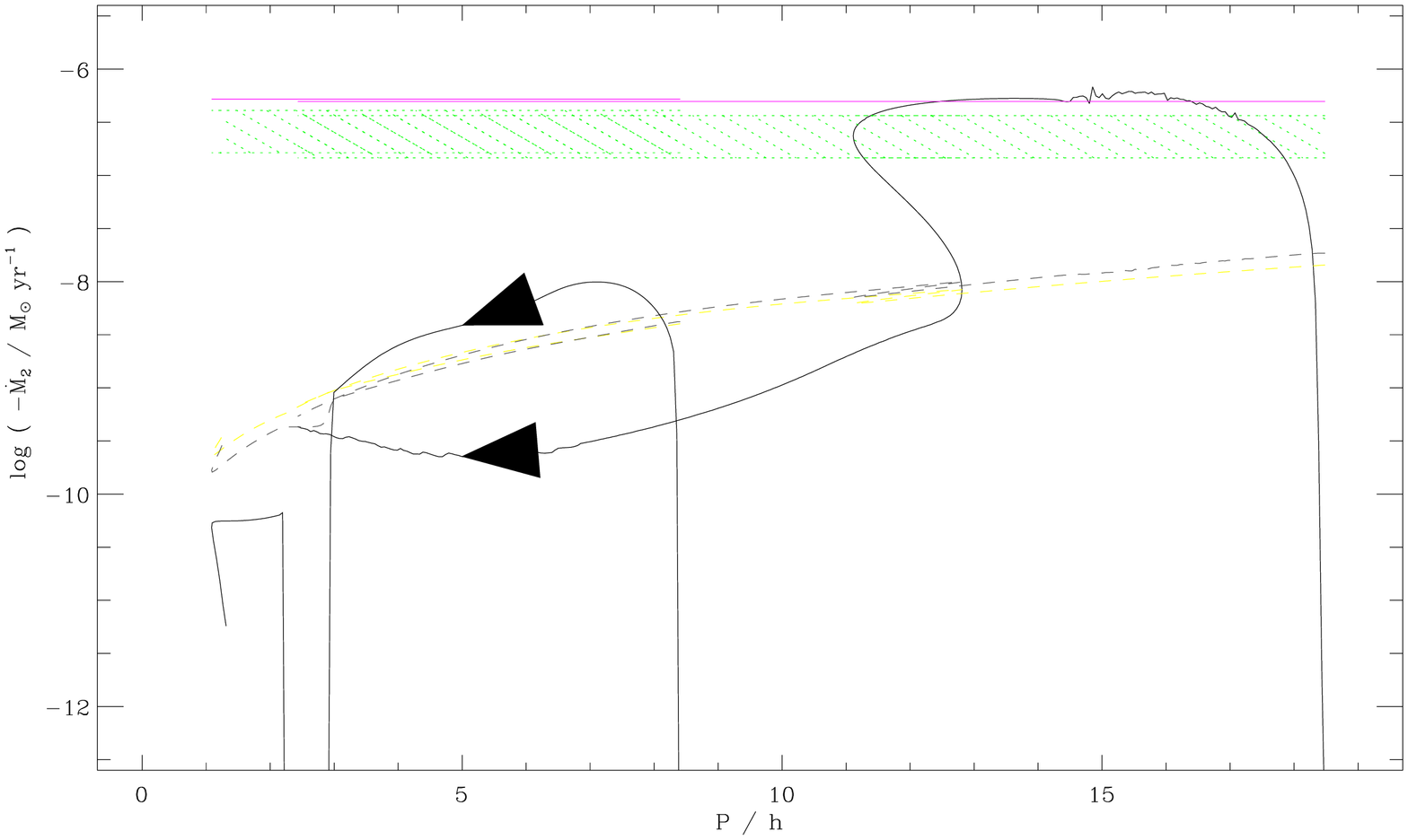, width=8cm, angle=90}
  \end{center}
\caption{ Mass transfer rate $-\dot M_2$ versus orbital period $P$ for
CVs. The curve starting near $P = 8.5$~h represents a standard CV
evolution under magnetic braking and gravitational radiation,
beginning with an unevolved secondary star. The mass transfer rate
falls to zero in the CV period gap $2~{\rm h} \la P \la 3~{\rm
h}$. The second $-\dot M_2(P)$ curve is for a slightly evolved
secondary initially more massive than the white dwarf. This undergoes
thermal--timescale mass transfer, including an episode of stable
nuclear burning of the matter accreting on to the white dwarf (shaded
band) before the mass ratio decreases sufficiently that the system
reaches the CV period range with a lower mass transfer rate than the
standard evolution. The dashed curves show the disc stability
criterion $T_{\rm visc}(R_{\rm out}) > T_{\rm H}$. These differ
because $R_{\rm out}$ is different for the two evolutions. Figure by
Klaus Schenker.  }
\label{cvfig}
\end{figure} 

Schenker et al. (2002) give an explicit example of this type of
evolution, modelling the CV AE~Aquarii. This system is not a dwarf
nova, but instead has a rapidly spinning magnetic white dwarf which
centrifugally expels most of the matter transferred to it. The fact
that the spin has not attained some kind of equilibrium in which
accretion is allowed strongly suggests that the mass transfer rate has
dropped very sharply in the recent past, on a timescale short compared
with the observed spindown timescale of $\sim 10^7$~yr. This is highly
suggestive of the ending of thermal--timescale mass transfer when the
binary mass ratio becomes sufficiently small; the current masses are
$M_1 \simeq 0.9\msun, M_2 \simeq 0.6\msun$, Welsh et al., 1995;
Casares et al., 1996. These masses themselves are suggestive: the
white dwarf may have accreted some matter from the companion, implying
steady nuclear burning and thus a thermal--timescale mass transfer
rate, while the secondary is clearly larger than its main--sequence
radius given the current binary period $P = 9.88$~hr (Welsh
et al., 1993). Finally AE Aqr's ultraviolet spectrum (Jameson et al.,
1980) shows extremely strong NV$\lambda$1238, but the usual
corresponding resonance line CIV$\lambda$1550 is completely
undetectable. As these lines have virtually identical ionization and
excitation conditions this is strongly suggestive of an abundance
anomaly in which nitrogen is enhanced at the expense of carbon. This
is turn is a signature of CNO processing, which clearly requires the
star to have been more massive ($\ga 1.5\msun$) in the past.

Schenker et al (2002) show that descent from thermal--timescale
evolution gives a consistent picture of AE~Aqr (see Figure
\ref{aefig}). This route therefore offers a way of introducing an
intrinsic spread in $-\dot M_2(P)$. Generally the mass transfer rates
are lower than for ZAMS secondaries, which may explain some of the
dwarf novae observed at periods above the gap (Fig \ref{cvfig}). In this
picture many dwarf novae should show signs of chemical
evolution. Evidently a good way of checking for this is to use the
ratio of the ultraviolet resonance lines of carbon and nitrogen
(CIV$\lambda$1550 to NV$\lambda$1238) as explained above.

Descendants of thermal timescale mass transfer are expected on
phase--space grounds to make up as much as 50\% of short--period CVs
(see Schenker et al, 2002) and so may account for the incidence of
dwarf novae at such periods. The main difficulty for this type of
explanation is that as seen from Fig. (\ref{cvfig}) there is a
tendency for CVs with evolved secondaries to fill the CV period gap. A
possible resolution of this difficulty is that the system moves very
rapidly through such periods, but more work is needed to substantiate
this.

\subsection{Long--period dwarf novae}

We have so far discussed only short--period ($P \la 12$~hr) CVs. There
is a small number of systems at longer periods, the best--studied
being GK Per ($P = 48$~hr). All of these systems appear to have
outbursts. This is exactly what we would expect from the stability
condition (\ref{16}). As the period of a CV increases, so does the
disc size $R_{\rm out}$, roughly as the binary separation $a \propto
P^{2/3}$. From (\ref{8}, \ref{16}) a stable disc then requires
\begin{equation}
-\dot M_2 > \dot M_{\rm crit} \propto R_{\rm out}^3 \propto P^2.
\label{dn}
\end{equation}
At such periods mass transfer is driven by nuclear expansion of the
secondary, at a rate
\begin{equation}
-\dot M_2 \propto P
\label{ev}
\end{equation}
roughly (see eq.\ref{mdotev} below). Hence at sufficiently long
periods all systems are likely to be dwarf novae. The coefficients in
(\ref{dn}, \ref{ev}) show that this will hold for $P \ga 1$~d, in
agreement with observation.

It is interesting to ask why relatively few CVs are seen at these
periods. There are several effects selecting against finding them, the
strongest probably being that they would only be found in outbursts,
which may be rather rare. Nevertheless these systems may be important,
as they may offer a channel for making Type Ia supernovae (King, Rolfe
\& Schenker, in prep.)


\section{Soft X--ray transients: the nature of the outbursts}
\label{sxtnat}

The comparative success of the disc instability idea in explaining
dwarf novae, and the qualitative similarities with soft X--ray
transients, make it natural to ask if the accretion discs in SXTs are
subject to the same instability. The major obstacle here is the vastly
different timescales noted earlier. One way of accomodating this
mathematically is simply to reduce the disc viscosity in SXTs compared
with dwarf novae. However there is no physical motivation for this
step, which cannot be regarded as plausible.

A more likely explanation of the long timescales in SXTs uses the
observed fact that the accretion discs in SXT outbursts (and indeed in
persistent LMXBs) are heavily irradiated. Van Paradijs \& McClintock
(1994) show that the optical brightness of outbursting SXTs and
persistent LMXBs correlates strongly with their X--ray luminosity, and
is far greater than would be expected from local viscous dissipation
within the disc (cf eqn \ref{8}). Irradiation thus raises the surface
temperature of the disc, and may potentially stabilize it by removing
its ionization zones. The early work on irradiated discs all followed
van Paradijs (1996) in using the
formula 
\begin{equation}
T_{\rm irr}(R)^4 = {\eta \dot M_{\rm c}c^2(1-\beta_a)\over 4\pi \sigma R^2}
\biggl({H \over R}\biggr)^n
\biggl[{{\rm d}\ln H\over {\rm d}\ln R} - 1\biggr], 
\label{irr}
\end{equation}
where $\eta$ is the efficiency of rest--mass energy conversion into
X--ray heating, $\dot M_{\rm c}$ is the central accretion rate, $H$
the disc scaleheight at disc radius $R$, $\beta_a$ is the albedo of
the disc faces, and the factor in square brackets lies between 1/8 and
2/7. The index $n = 1$ or 2 depending on whether there is a central
irradiating point source or not; this is discussed further below. The ratio
$H/R $ is roughly constant in a disc, so $T_{\rm irr}$ falls off as
$R^{-1/2}$. Thus for a large enough disc, $T_{\rm irr}$ dominates the
disc's own effective temperature $T_{\rm visc}$, which goes as
$R^{-3/4}$. This agrees with our expectations above.

Constructing a self--consistent irradiated disc is a difficult
theoretical problem, and attempts to date do not produce results in
agreement with observation. In particular calculations of axisymmetric
discs with irradiation (e.g. Kim et al., 1999; Dubus et al., 1999;
Tuchman et al., 1990) tend to show that the inner parts of the disc
expand and shadow the outer disc from the irradiation. As most of the
disc mass is at large radii, this effect would prevent irradiation
having a significant effect on the outbursts. However since there is
abundant observational evidence that LMXB discs {\it are} strongly
irradiated, it is safe to assume this; we shall return later to the
question of why axisymmetric disc calculations have difficulty in
reproducing this result.

Now let us consider an outburst in a soft X--ray transient. This will
be triggered by the ionization instability at some disc radius, and
eventually lead to matter accreting strongly on to the black hole or
neutron star at the disc centre. At this point the X--ray outburst
begins: the central X--ray emission irradiates the disc, and heats
it. If the disc is small enough the whole of it will now have a
surface temperature $T_{\rm irr}$ above $T_{\rm H}$ and be in the hot,
high--viscosity state. This reinforces the tendency of mass to accrete
inwards, and reduces the local surface density $\Sigma$ on a viscous
timescale.  If there were no irradiation, as in a dwarf nova, $\Sigma$
would eventually at some radius reach the value $\Sigma_{\rm min}$
where the disc must jump back (on a thermal timescale) to the cool,
low--viscosity state. This in turn would trigger a cooling wave to
move across the disc and return it all to the low state, ending the
outburst. However if the disc is irradiated by the central X--ray
source this cannot happen: the local temperature is fixed not by local
viscous dissipation, but by irradiation. As this is fuelled by the
central accretion rate it is a globally rather than locally determined
quantity. The disc is trapped in the hot state everywhere until the
central accretion rate declines to the point where irradiation can no
longer keep the disc in this state. But if there is no cooling wave,
the central accretion rate can only decline as a result of the
accretion of a significant fraction of the disc mass. It follows that
irradiation is likely to prolong disc outbursts and make them use up
more of the disc mass.  For a given mass transfer rate, the latter
effect will lengthen the quiescent intervals in which the disc mass is
rebuilt by accretion from the companion star. 

This line of reasoning allowed King \& Ritter (1998) to give a simple
explanation of why SXTs have much longer outburst and quiescent
timescales. Their treatment also shows that the X--ray light curves
are likely to have certain characteristic shapes. An outbursting disc is
approximately in a steady state with surface density
\begin{equation}
\Sigma(R) \simeq {\dot M_{\rm c}\over 3\pi\nu}
\end{equation}
\label{eq:Sig}
where $\dot M_{\rm c}$ is the central accretion rate.
Integrating this gives the total initial mass of the hot zone as 
\begin{equation}
M_{\rm h} = 2\pi\int_0^{R_{\rm h}}\Sigma R{\rm d}R \simeq 
                                 \dot M_{\rm c}{R_{\rm h}^2\over 3\nu} 
\label{eq:mh}
\end{equation}
since the inner disc radius is much smaller than the outer radius
$R_{\rm h}$ reached by the heating front. In (\ref{eq:mh}) $\nu$ is
some suitable average of the kinematic viscosity in the disc, close to
its value near $R_h$. Note that (\ref{eq:mh}) is effectively a
dimensional relation, and simply asserts the obvious fact that the
mass of a steady disc is given by the product of the accretion rate
and the viscous time at its outer edge: it does not for example assume
that $\nu$ is constant through the disc.

As we reasoned above, the only way in which
the mass of the hot zone can change is through central accretion, so we
have $\dot M_{\rm c} = -\dot M_{\rm h}$, and
\begin{equation}
-\dot M_{\rm h} = {3\nu\over R_{\rm h}^2}M_h
\label{eq:dotm}
\end{equation}
or
\begin{equation}
M_{\rm h} = M_0e^{-3\nu t/R_h^2},
\label{eq:mh2}
\end{equation}
where $M_0$ is the initial mass of the hot zone. This in turn implies
that the central accretion rate, and thus the X--ray emission, decays
exponentially, i.e.
\begin{equation}
\dot M_c = {R_h^2M_0\over 3\nu}e^{-3\nu t/R_h^2}.
\label{18}
\end{equation}
Note that the peak accretion rate at the start of the outburst can be
expressed as (disc mass)/(viscous time of the entire hot disc). If the
quiescent disc is close to the maximum mass allowed before becoming
unstable, this peak rate depends only on the disc size: King \& Ritter
(1998) find
\begin{equation}
\dot M_{\rm c}({\rm peak}) \simeq 
4.8\times10^{-8}R_{11}^{7/4}\ M_{\odot}\ {\rm yr}^{-1}
\label{eq:mob}
\end{equation}
for small (fully irradiated) discs, and
\begin{equation} 
\dot M_{\rm c}({\rm peak}) \simeq 4.1\times 10^{-8}R_{12}^2\ M_{\odot}\
{\rm yr}^{-1}
\label{eq:mob2}
\end{equation}
for large discs, where $R_{11}, R_{12}$ are the disc radii in units of
$10^{11}$~cm and $10^{12}$~cm respectively. These expressions agree
with the fact that most SXT outbursts are observed to be close to the
Eddington luminosity, even for cases where the accretor is a $\sim
10\msun$ black hole. The decay constant for
outbursts is
\begin{equation}
\tau \simeq 40R_{11}^{5/4}\ {\rm d} \simeq 2R_{12}^{5/4}~{\rm yr},
\label{eq:tau}
\end{equation}
showing that outbursts can be very prolonged in wide systems.

Eventually $\dot M_{\rm c}$ drops to the point that irradiation cannot
keep the outer edge of the disc in the hot state.  The outburst
proceeds nevertheless, as the central irradiation keeps the inner disc
regions ionized. A simple calculation shows that the X--rays then
decay linearly rather than exponentially, still on the hot--state
viscous timescale. A sufficiently large disc cannot be kept in the hot
state by irradiation even at the start of the outburst, and so is
always in the linear regime.

A more exact treatment (King, 1998) solves the diffusion equation
(\ref{4}) for an irradiated disc.  The solutions predict steep
power--law X--ray decays $L_X \sim (1+ t/t_{\rm visc})^{-4}$, changing
to $L_X \sim (1-t/t'_{\rm visc})^4$ at late times, where $t_{\rm
visc}, t'_{\rm visc}$ are viscous timescales. These forms closely
resemble the approximate exponential and linear decays inferred above
in these two regimes.  It is important to realise that the decays are
quite different than for unirradiated discs because the viscosity is a
function of the central accretion rate rather than of local conditions
in the disc.

Since disc size scales with the binary separation, and thus as
$P^{2/3}$ by Kepler's law, we arrive at a simple picture in which SXT
outbursts in short--period systems begin as exponential, becoming
linear near the end of the outburst. In long--period systems
the outbursts may be linear throughout. These basic features are
generally, but not universally, found in observations. Thus
short--period systems such as A0620--00 ($P = 7.8$~hr) have classic
`FRED' (fast rise, exponential decay) light curves (see Ch ** of this
book), whereas GRO~J1744--28 (with $P = 11.8$~d one of the
longest--period SXTs) has an entirely linear decay (Giles et al.,
1996). The exponential--linear dichotomy is examined in detail by
Shahbaz et al. (1998), and is in good agreement with observation. 

While this simple picture is largely correct, there are a number of
complications which we should address. We return first to the problem
mentioned above, namely that axisymmetric calculations of irradiated
discs tend to produce configurations which are strongly
self--shadowed, and thus very unlike what is observed. The resolution
of this difficulty comes from the realisation (Pringle, 1996) that
as discussed in Section 2.5 above,
strong self--irradiation causes an accretion disc to warp in a
non--axisymmetric fashion, with irradiation possible at all
radii. Even though this irradiation is relatively patchy (see Fig. 7
of Pringle, 1997) it is likely to keep the disc ionized if it is
intense enough, i.e. if $T_{\rm irr} > T_{\rm H}$, because recombination
times are long compared with the dynamical time. This suggests a
reason why irradiation seems to be unreasonably effective, as
discussed above. A second consequence of warping will be very
important later on, namely that the radiation field must become quite
anisotropic. This follows from considering the disc angular momentum
vectors ${\bf j}(R)$ measured wrt the accreting object. An unwarped
disc has all the ${\bf j}(R)$ vectors parallel to that of the binary,
${\bf J}_{\rm orb}$. However, the ${\bf j}(R)$ are clearly scrambled
in many directions once warping has taken place. Since matter joined
the disc with ${\bf j}(R)$ parallel to ${\bf J}_{\rm orb}$, the
radiation field must have changed its own (originally zero) angular
momentum to compensate. This must mean that it is anisotropic, and
indeed this is what Pringle's calculations reveal (Pringle, 1997;
Fig. 7). In general the radiation field at infinity is confined to a
fairly narrow double cone.

\begin{figure}
  \begin{center}
    \epsfig{file=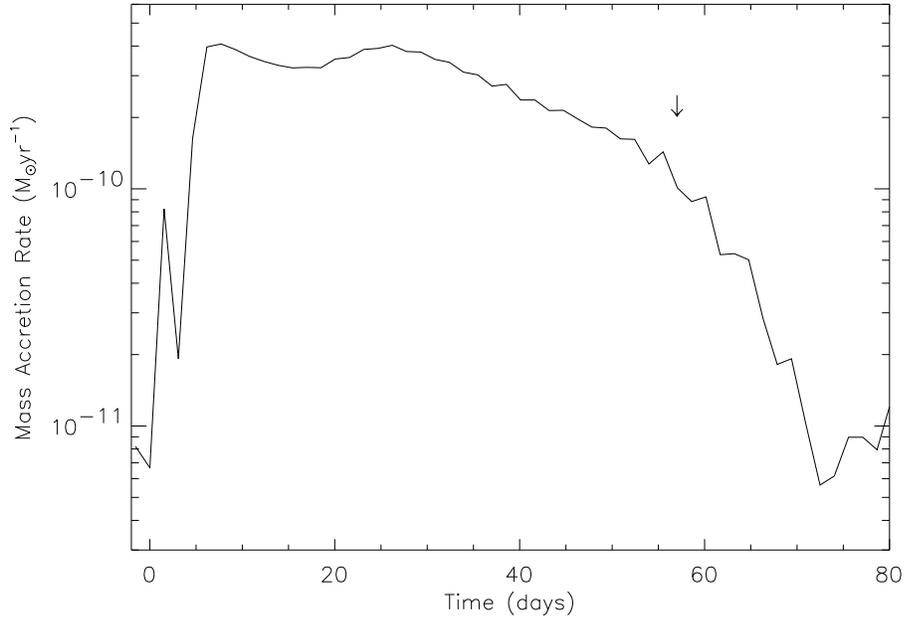, width=12cm}
  \end{center}
\caption{ Simulated outburst of a soft X--ray transient showing a
secondary maximum, here after $\sim 26$~d (Truss et al.,
2002). Note that the slopes of the light curve are similar each side
of this maximum.}
\label{sxtfig}
\end{figure}

In addition to the exponential or linear X--ray light curves discussed
above, many SXT outbursts show a secondary maximum (factors of a few)
in their lightcurves once the X--rays have declined by a few
e-folds. This increase signals the accretion of a new source of mass,
which is rather smaller than the original heated disc region giving
rise to the basic light curve shape. There have been several attempted
explanations of this, sometimes invoking extra mass transfer from the
secondary star, but a recent 2--D simulation of SXT outbursts
(Truss et al., 2002) reveals a likely cause. The simulations confirm
the simple picture described above as the cause of the main
outburst. The secondary maximum results from two effects not included
in the arguments above. First, even in a disc irradiated to its outer
edge, some of the matter at the outer disc rim will remain in the cool
state, because the outer edge of the disc flares and shields
it. Second, the mass ratios in all SXTs allow the disc to reach the
3:1 resonant radius. This is fairly obvious for black--hole systems,
where $M_1 \sim 5 - 10\msun$, but we will see shortly that it is true
of neutron--star transients also. 

The result of including these two effects in the simulations is to
produce a second, superoutburst--like increase in the accretion rate
through the disc (Fig. \ref{sxtfig}), aided by the strong tidal torque
at the resonant radius.

The main effect causing a deviation from linear decays in long--period
systems is that by definition, such systems can have a large fraction
of their disc mass permanently in the low state. This mass reservoir
can give rise to hysteresis effects, distorting the simple picture
predicting linear decays. The mass reservoir can also allow bursts to
recur more frequently than might be expected. For example,
GRO~J1744--28 was completely undetected in $\sim 30$~yr of X--ray
astronomy until its first outburst, but then had another outburst only
a few months later. For very wide systems the outbursts can also be
very long, and indeed some LMXBs usually classified as `persistent'
may actually be in outbursts which have lasted for the entire history
of X--ray observations. Proof of this comes from the longest--period
SXT currently known, GRS~1915+105, which again was not detected until
it went into outburst in 1992. With some variability it has remained
bright ever since, showing that outbursts in wide systems can
certainly last for at least a decade. The binary period of 33.5~d and
mass $\sim 15\msun$ (Greiner et al., 2001) show that the disc radius
here must be $\sim 4\times 10^{12}$~cm, implying from eq.
(\ref{discmass}) below that $M_{\rm disc,\ max} \simeq 7\times
10^{29}$~g. Such a disc could supply the inferred outburst accretion
rate $\sim 10^{19}$~g~s$^{-1}$ for more than $10^3$~yr. This is far
longer than the decay constant $\tau \sim 10$~yr
(eq. \ref{eq:tau}). GRS~1915+105 provides an explicit example of the
complex behaviour of long--period SXTs, as it is observed to vary on
the timescale $\tau$ but maintains its outburst for longer.

An obvious corollary of these ideas is that there must exist a large
unseen population of quiescent transients which have never been
observed to have an outburst. In the next subsection we will see that
there is strong evidence for the existence of such objects. Taken
together, all these effects show that there is a general tendency for
outbursts in long--period systems to show a more complex variety of
behaviours than in short--period systems.

\begin{figure}
  \begin{center}
    \epsfig{file=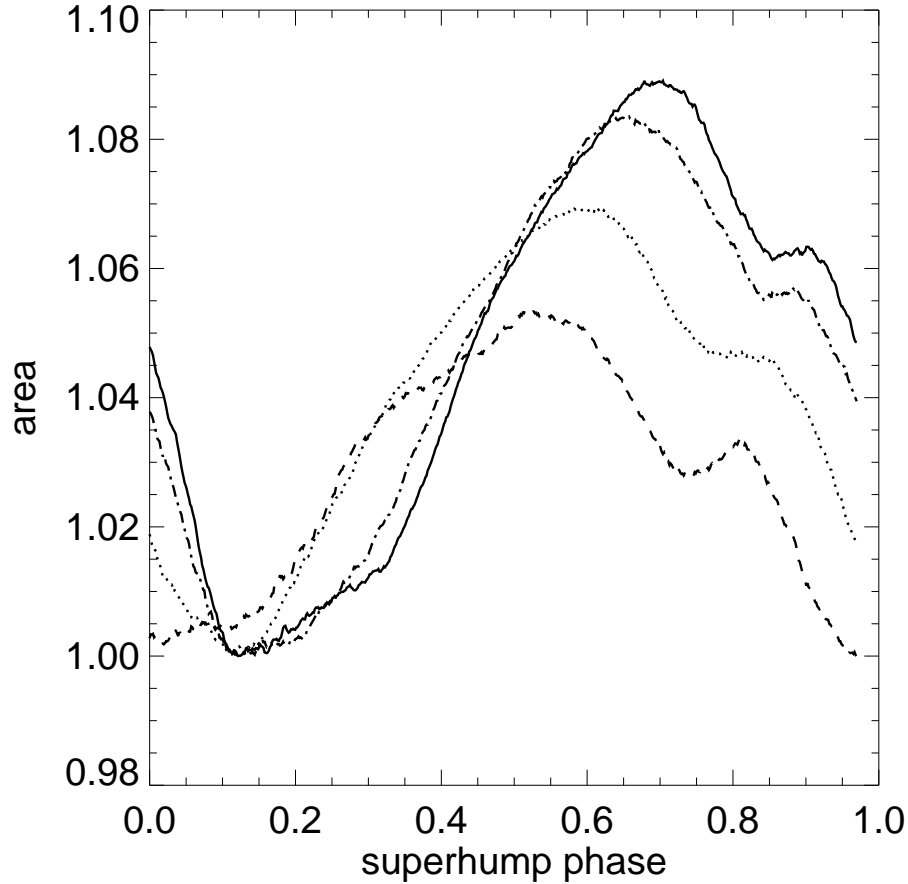, width=12cm}
  \end{center}
\caption{ Disc surface area as it varies over the course of a single
superhump cycle (Haswell et al., 2001). The solid line shows the
change in surface area of the entire disc (+ stream), the dot--dash
line shows disc region with densities $> 1\%$ of the maximum density,
dotted line shows regions with density $> 5\%$ of the maximum and the
dashed line shows regions with density $> 10\%$ of the maximum.  In
this simulation the superhump period is $1.0295 P_{\rm orb}$ .  }
\label{shfig}
\end{figure}

As mentioned above, all SXTs have mass ratios allowing the disc to
reach the 3:1 resonant radius, and thus one might expect superhumps to
appear. A survey of the optical data on these systems (O'Donoghue \&
Charles, 1996) concluded that indeed superhumps are observed in them.
However at first sight this seems paradoxical, as we have asserted
above that superhumps are a modulation of the viscous dissipation in
precessing discs, but also that irradiation completely outweighs this
dissipation in LMXB (including SXT) discs. The resolution of this
problem is that the superhump modulates not only the disc dissipation,
but also the disc area (Haswell et al., 2001). Since a larger area
intercepts and reradiates a large fraction of the central X--rays,
there should indeed be a superhump modulation of the optical light. In
fact for constant X--ray luminosity the optical light curve should
simply map the variation of the disc area (Fig. \ref{shfig}).
The resulting curve agrees well with observation. In general the
optical light curve is the convolution of the X--ray light curve with
this area variation.

\section{Soft X--ray transients: the occurrence of the outbursts}
\label{occob}

We should next consider the question of when SXT outbursts occur. Here
there have been major advances since the last edition of this book. It
is now clear that outbursts are extremely prevalent among LMXBs, so
that if anything, persistent systems are rather the exception. This
realization has in turn had important consequences for our
understanding of compact binary evolution. 

These advances stem from van Paradijs's (1996) realisation that the
correct condition for LMXBs to be stable against outbursts (and thus
appear as persistent systems) is
\begin{equation}
T_{\rm irr}(R_{\rm out}) > T_{\rm H} 
\label{sxt}
\end{equation}
rather than (\ref{16}). In his paper van Paradijs (1996) used the 
observed X--ray luminosities $L_{\rm X}$ to replace the combination
$\eta\dot M_c c^2$ in the expression (\ref{irr}) for $T_{\rm irr}$
(with $n=1$, appropriate for a strong central irradiating source).
He was able to demonstrate that indeed the condition (\ref{sxt})
correctly divides persistent LMXBs from transient systems.

\begin{figure}
  \begin{center}
    \epsfig{file=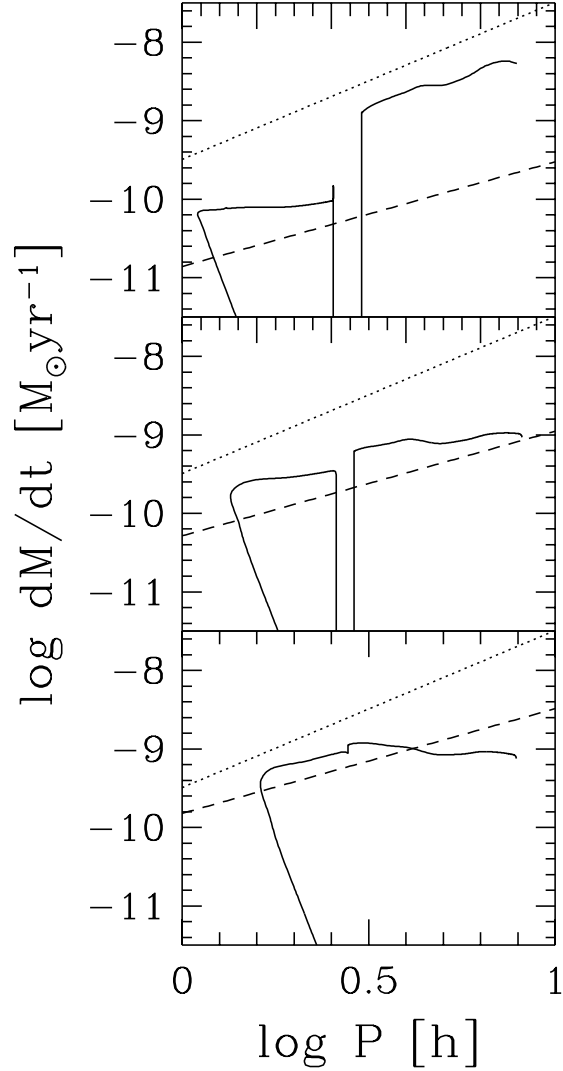, width=12cm}
  \end{center}
\caption{ Mass transfer rate versus orbital period for LMXBs with
various primary masses and an unevolved secondary of initial mass
$1\msun$ (King et al., 1996). The evolution is driven by magnetic
braking as long as the secondary has a convective core. The dotted and
dashed lines are the critical mass transfer rates given by (\ref{16})
and (\ref{sxt}) for standard and irradiated discs
respectively. Systems with lower mass transfer rates are
transient. {\it Top panel:} $1.4\msun$ neutron star primary. {\it
Middle panel:} $10\msun$ BH primary. {\it Bottom panel}: $50\msun$
primary. All systems are stable according to the irradiated-disc
criterion (\ref{sxt}). This shows that short--period transients must
have evolved secondaries.
}
\label{ulifig}
\end{figure}

This success means that we can now use the condition (\ref{sxt}) with
$T_{\rm irr}$ calculated from the evolutionary mean mass transfer rate
rather than the observed $L_x$, i.e. with $\dot M_c$ replaced by
$-\dot M_2$. The results are revealing.

\subsection{Short--period SXTs}
\label{sp}

For short orbital periods $P \la 12$~hr mass transfer must
be driven by angular momentum loss, and we might expect the secondary
stars to be unevolved low--mass main--sequence stars. Now Fig. (\ref{ulifig})
shows that if the secondaries in either neutron--star or black--hole
LMXBs are unevolved main--sequence stars, all systems are predicted to
be persistent. But observation (see Ch. **) shows that most, if not
all, short--period black--hole LMXBs are {\it transient}, and there
are also a number of neutron--star SXTs at such periods.

One possible way of avoiding this conclusion, at least for black-hole
systems, is the idea that the irradiation effect might be weaker for a
black--hole accretor. Shakura \& Sunyaev (1973) pointed out that if
there is no strong central source, the irradiation comes only from
central disc regions lying in the orbital plane, reducing its effect
by a second projection angle $\sim H/R$. Thus the value $n=2$ would be
appropriate in (\ref{irr}). King et al. (1997) found that indeed this
would make black--hole LMXBs transient even with unevolved secondary
stars. However the X--ray spectra of most black--hole systems have
a strong power--law component, which is usually thought to come from a
corona. In this case it is unlikely that the weaker irradiation law
with $n=2$ in (\ref{irr}) is appropriate for determining disc
stability.

\begin{figure}
  \begin{center}
    \epsfig{file=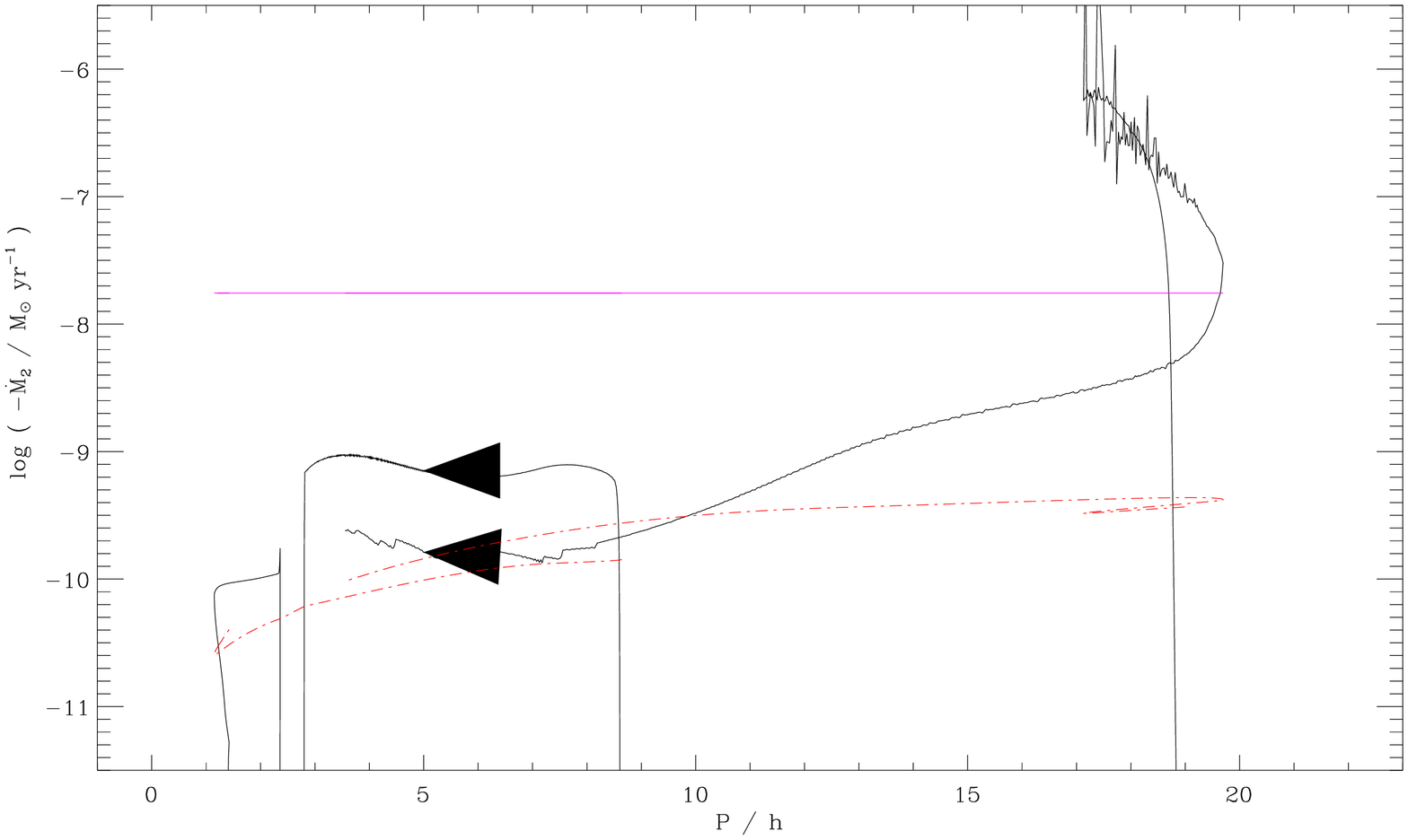, width=8cm, angle=90}
  \end{center}
\caption{ Mass transfer rate $-\dot M_2$ versus orbital period $P$ for
neutron--star LMXBs ($M_1 = 1.4\msun$). The curve starting near $P =
8.5$~h represents evolution under magnetic braking and gravitational
radiation, beginning with an unevolved secondary star. The mass
transfer rate falls to zero in a period gap $2.3~{\rm h} \la P \la
2.8~{\rm h}$. The second $-\dot M_2(P)$ curve begins with a slightly
evolved secondary, initially more massive ($2.35\msun$) than the
neutron star. This undergoes thermal--timescale mass transfer (the
horizontal line shows the Eddington limit) before reaching short
periods with a lower mass transfer rate than the standard
evolution. The dashed curves show the disc stability criterion $T_{\rm
irr}(R_{\rm out}) > T_{\rm H}$. These differ as the two evolutions have
different $R_{\rm out}$. The unevolved system is always persistent,
while the system with the evolved secondary is transient in the period
range 5 -- 10~hr. Compare Fig. (\ref{cvfig}). Figure by Klaus Schenker.
}
\label{nssxtfig}
\end{figure}

\begin{figure}
  \begin{center}
    \epsfig{file=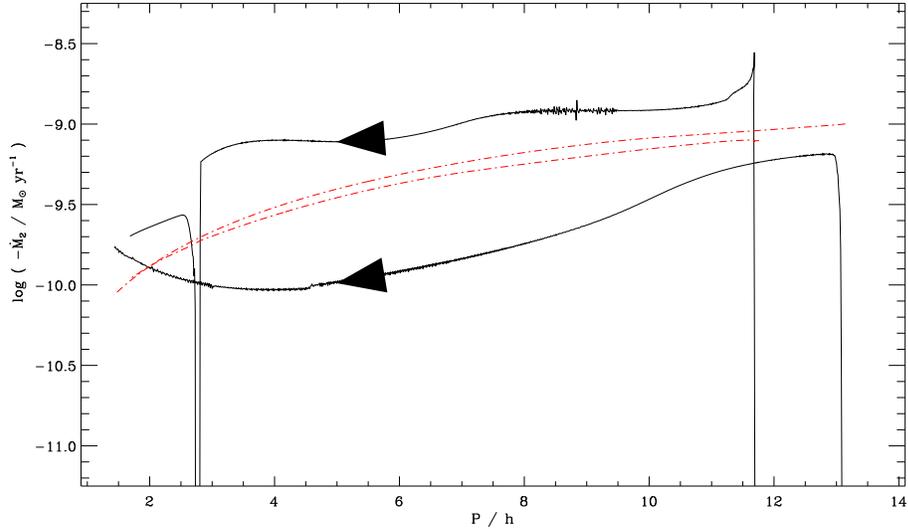, width=8cm, angle=90}
  \end{center}

\caption{As for Figure \ref{nssxtfig}, but for black--hole LMXBs ($M_1 =
7\msun$).The curve starting near $P = 11.8$~h represents evolution
under magnetic braking and gravitational radiation, beginning with an
unevolved secondary star of mass $1.5\msun$. The curve starting near
$P = 13$~h shows the evolution starting with a significantly
nuclear--evolved secondary of $1\msun$. As before the dashed curves
are the stability criterion $T_{\rm irr}(R_{\rm out}) > T_{\rm H}$ in
the two cases. Again only the system with the evolved secondary is
transient. Figure by Klaus Schenker.  }
\label{bhsxtfig}
\end{figure}

\begin{figure}
  \begin{center}
    \epsfig{file=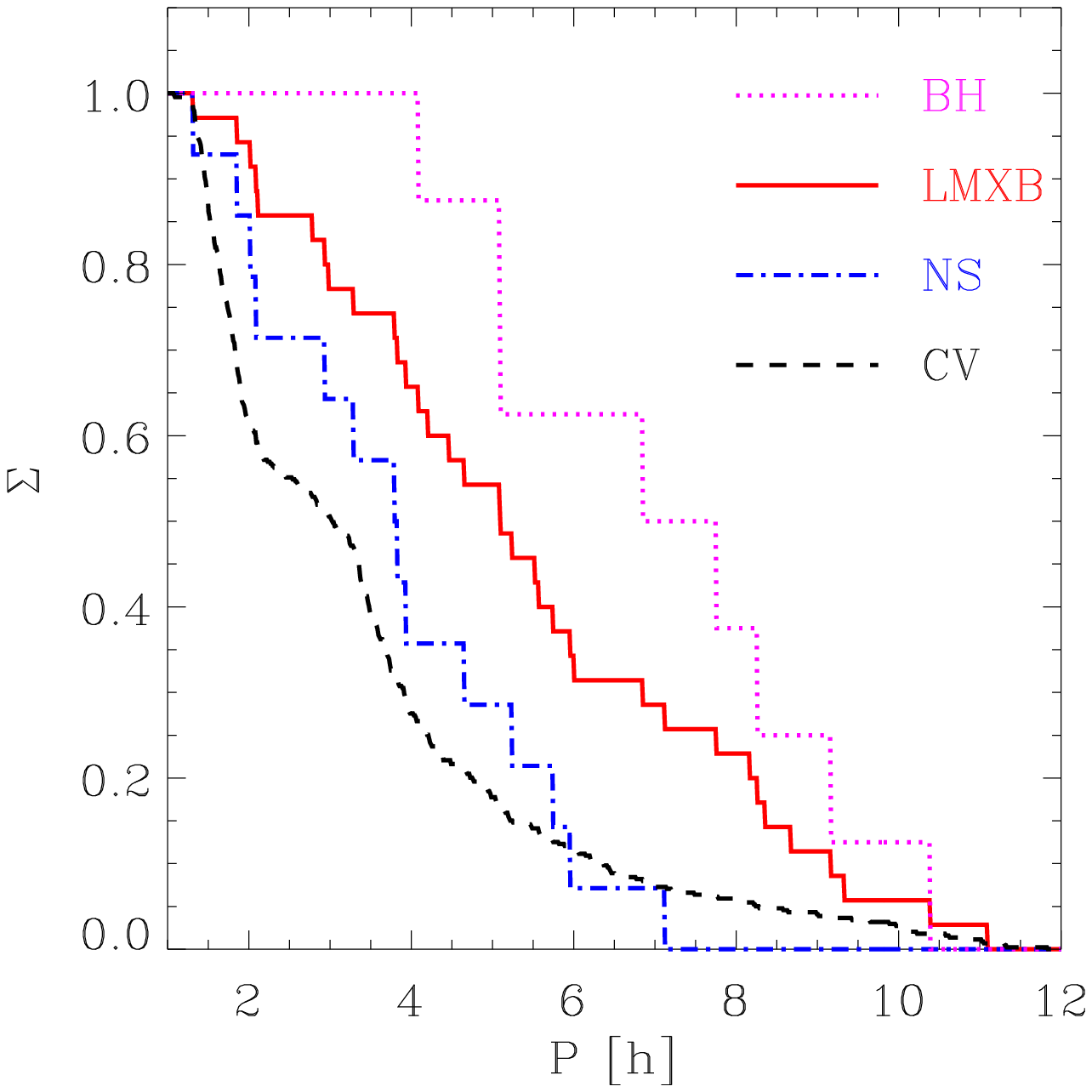, width=12cm, angle=0}
  \end{center}

\caption{The cumulative period histogram below $P =12$~hr for LMXBs
(solid) compared with that for CVs (dashed). Also shown are the
histograms for black hole LMXBs (dotted) and neutron--star LMXBs
(dot--dashed). The CV histogram clearly shows a flatter slope between
period $P = 3$ and 2 hours, corresponding to the well--known period
gap. There is no significant evidence for such a feature in the LMXB
histogram, which is consistent with a uniform
distribution. Neutron--star and black--hole systems appear to have
significantly different period distributions. Data from Ritter \&
Kolb, 2003. Figure by Klaus Schenker.  }
\label{perioddist}
\end{figure}

If this explanation is abandoned, it seems that the inevitable
conclusion (King et al., 1996) is that despite appearances, the
secondary stars are actually chemically evolved in a large fraction of
short--period LMXBs -- if not a majority or even a totality in the
black--hole case. This must mean that they descend from stars with
initial masses greater than $0.8\msun$, and have had time to evolve
away from the ZAMS before mass mass transfer driven by angular
momentum loss has pulled them in to short orbital periods. Figures
\ref{nssxtfig} and \ref{bhsxtfig} show explicitly that this kind of
evolution does produce short--period transient systems.

\begin{figure}
  \begin{center}
    \epsfig{file=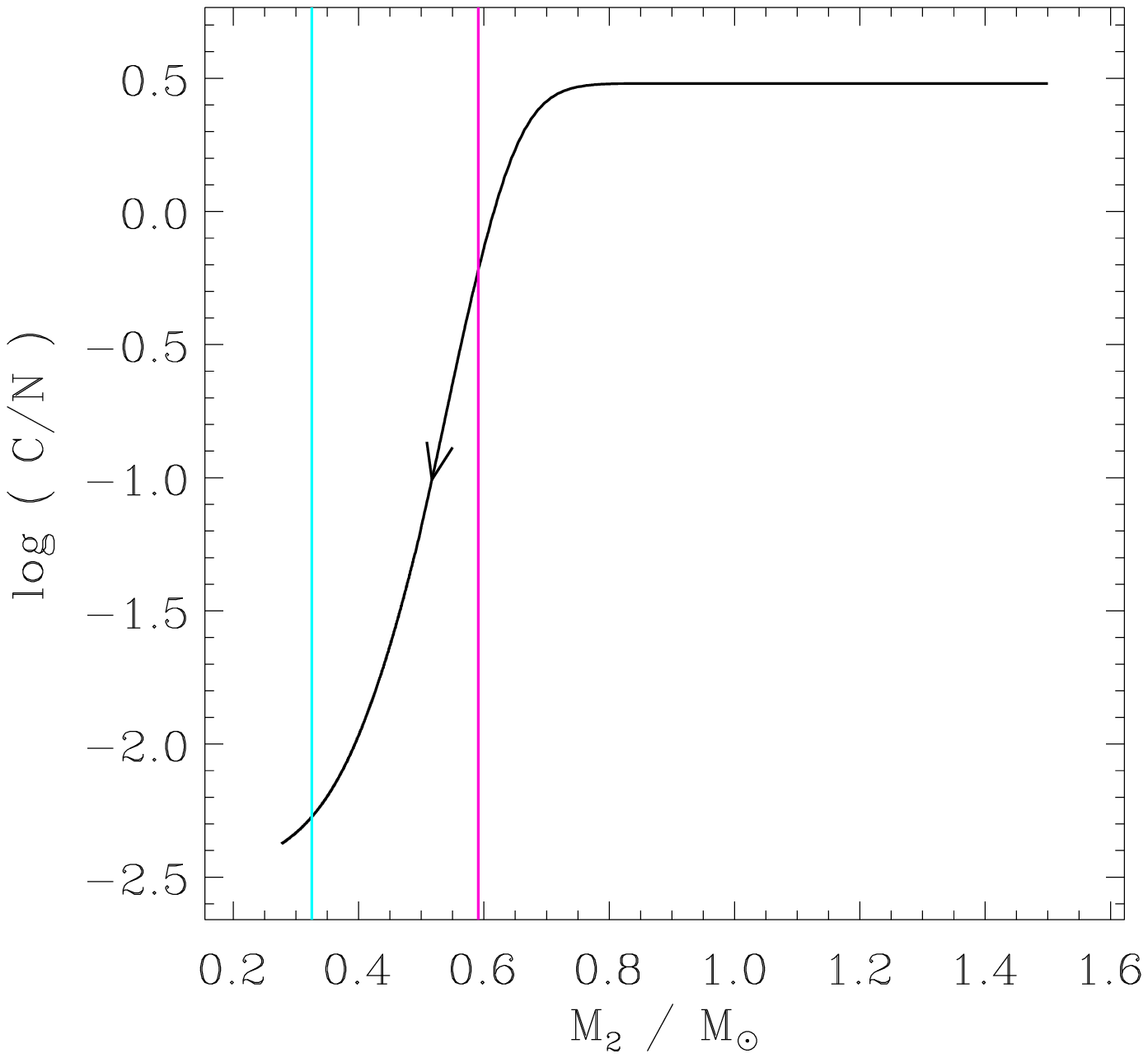, width=12cm}
  \end{center}
\caption{Evolution of the surface abundance ratio C/N for a binary
beginning mass transfer from a $1.5 \, \msun$ main sequence star on to
a $7 \, \msun$ black hole. The C/N ratio remains at the cosmic value
($\log(C/N) \simeq 0.5$) until the companion mass $M_2$ is reduced
below $0.8\msun$. Until this point chemical changes are confined to
the interior: at the turn--on period of 15~hr the core hydrogen
fraction had already been reduced to 28 \%. The current period of
XTE~J1118+480 is reached at a mass of $0.33 \, \msun$, indicated by
the left vertical line, while the other near $0.6 \, \msun$ shows the
period of A0620-00. The transferred mass has been accreted by the
black hole which has grown beyond $8 \, \msun$. Figure from Haswell et
al., 2002.)}
\label{1118fig}
\end{figure}

This shift of view parallels the similar shift in our view of CVs (see
subsection (1.4.1) above), where we now believe that there is a
significant admixture ($\la 50\%$) of evolved secondaries in CVs which
have descended via the thermal--timescale mass transfer route. But the
shift for short--period LMXBs may be more drastic: it could be that
{\it all} black--hole LMXBs have evolved secondaries, as so far there
is no convincing evidence that {\it any} black--hole LMXB has an
unevolved secondary. Black--hole systems with main--sequence
companions would appear among the handful of persistent LMXBs which do
not have Type I X--ray bursts (signalling the presence of a neutron
star). They are clearly difficult to identify, as in persistent
systems one cannot measure a mass function and so get a dynamical mass
estimate. However we will see that even some persistent LMXBs probably
have evolved secondaries.

An independent line of argument leading to a similar conclusion comes
from the period distribution of LMXBs (Fig \ref{perioddist}). This
shows no sign of the familiar CV period gap between 2 and 3 hours
orbital period. This suggests that in LMXBs accretion either does not
cease at 3~hr, or does so in a narrower period range, which may also
differ for individual systems. This is just what we see in
Figs. (\ref{nssxtfig}) and (\ref{bhsxtfig}), showing the mass transfer
rates for neutron--star and black--hole LMXBs which have reached short
periods with secondaries having some degree of nuclear evolution.

Confirmation of these ideas comes from observations of the
short--period black--hole SXT XTE~J1118+480 (Haswell et al.,
2002). The Hubble Space Telescope ultraviolet spectrum of this object
shows extremely strong NV$\lambda$1238, but CIV$\lambda$1550 is
completely undetectable. As with AE~Aqr (see above) this is a clear
sign of CNO processing. The secondary star must have been more massive
($\ga 1.5\msun$) in the past than the present binary period (4.1~hr)
would suggest. It was evidently somewhat chemically evolved when mass
transfer began. For much of its life this would not have been obvious
from the composition of its outer layers and thus of the mass visible
in emission lines: Fig. (\ref{1118fig}) shows that this would have
appeared completely normal until the secondary mass got low enough
for convection to mix the processed layers into the envelope.

\begin{figure}
  \begin{center}
    \epsfig{file=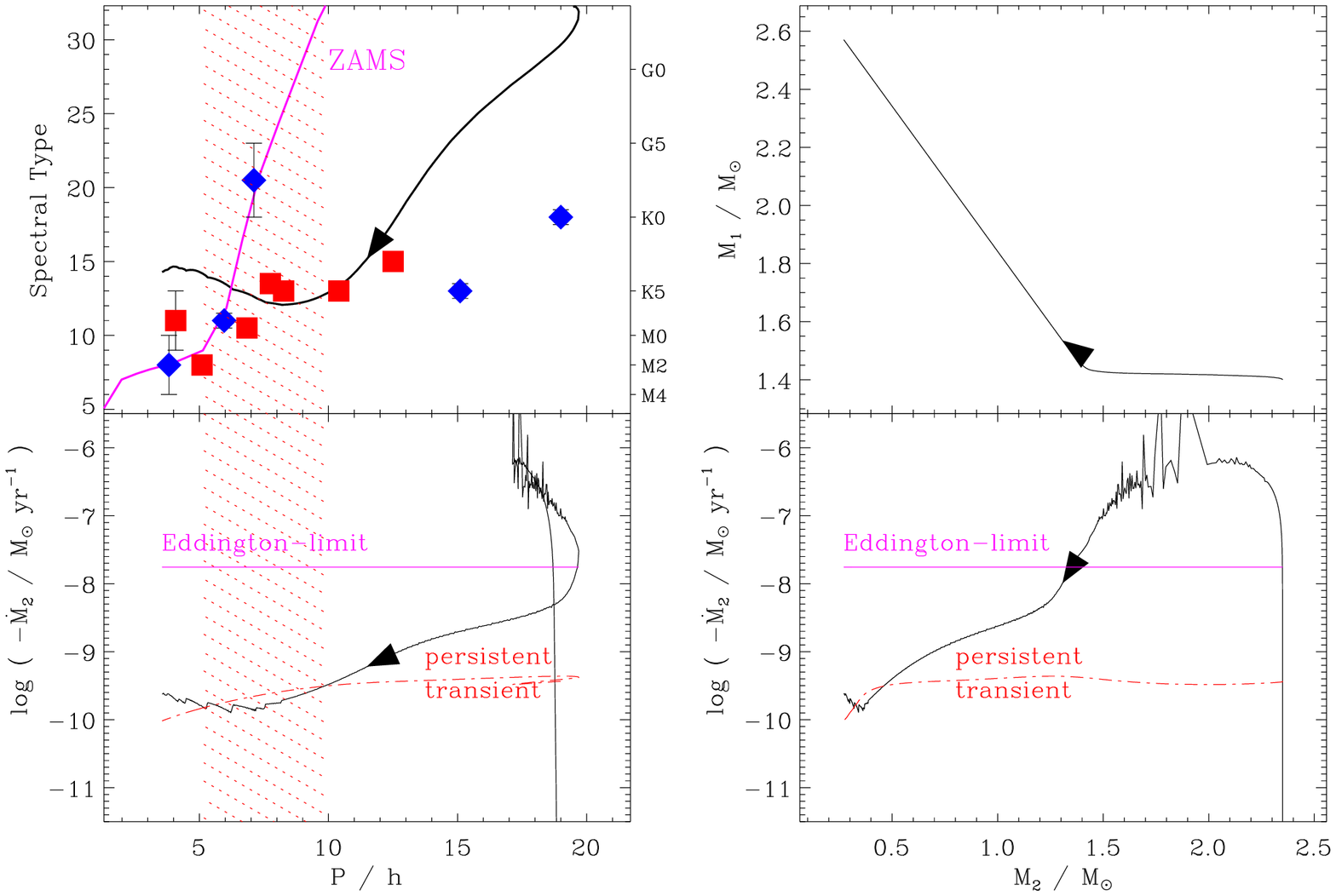, width=8cm, angle=90}
  \end{center}
\caption{The evolution of short--period neutron--star LMXBs. The
initial companion mass is $2.35\msun$, with core hydrogen fraction
35\%. The system first goes through a phase of thermal--timescale
mass transfer ($M_2 \ga 1.5\msun$ in lower rh panel, $P \sim 17 -
20$~hr in lower lh panel). After this phase the mass transfer rate
drops below the Eddington limit and the neutron star begns to grow
significantly in mass (upper rh panel). At periods between 10 and 5~hr
the system becomes transient (shaded region in lh panels). The
spectral type of the secondary (solid curve in top lh panel is close
to that which a ZAMS secondary would have (light curve) at these
periods. Varying the mass and initial degree of evolution of the
secondary at the onset of mass transfer produces a family of curves
displaced horizontally from the solid curve. The squares show
persistent neutron--star LMXBs and the diamonds are neutron--star
transients.}
\label{nslmxbfig}
\end{figure}
\begin{figure}
  \begin{center}
    \epsfig{file=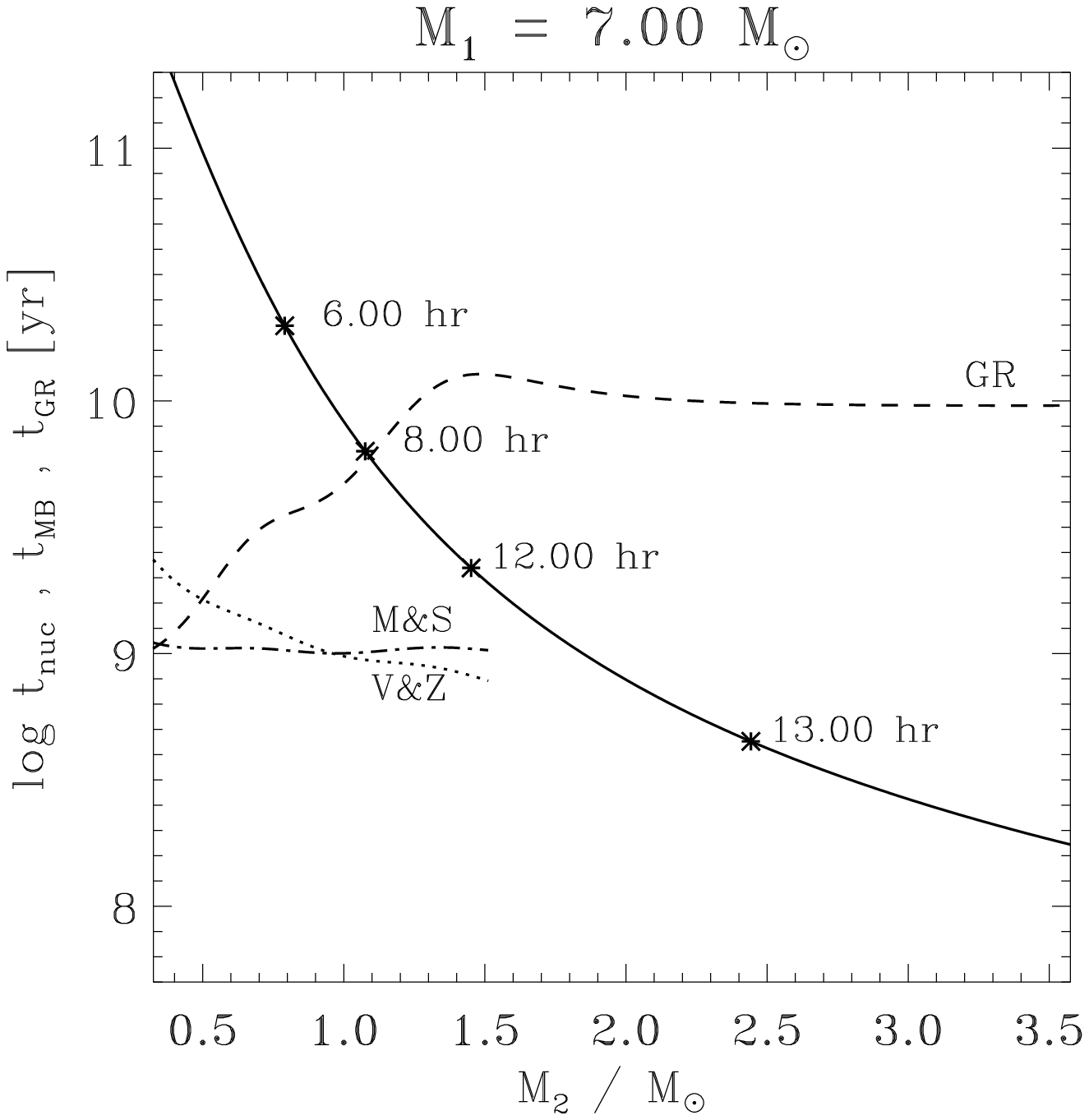, width=12cm}
  \end{center}
\caption{ Comparison of timescales for a $7 \, \msun$ black hole
primary.  The full curve gives the nuclear timescale as a function of
the secondary mass.  At various points the orbital period for a ZAMS
star filling its Roche lobe is indicated.  The other 3 curves show the
various relevant angular momentum loss timescales ($(-{\rm d} \ln
J/{\rm d} t)^{-1}$): for gravitational radiation (GR, dashed) and
magnetic braking according to Mestel \& Spruit (M\&S, dash-dotted) and
Verbunt \& Zwaan (V\&Z, dash-triple-dotted), all in the version of
Kolb (1992) Magnetic braking is assumed to be quenched in stars which
have no convective envelopes, i.e for $M_2 \ga 1.5\msun$.
}
\label{taunucfig}
\end{figure}

Given the pervasive evidence that short--period SXTs have evolved
companions we should now ask why this is so. In the black--hole case
the answer seems to go back to evolution of the evolution of the
binary as the black--hole progenitor expands off the main sequence.
There are two cases. The binary may be so wide at this stage that the
companion never interacts with the primary's envelope as this
expansion occurs. Such systems clearly do not reach contact and
produce a BHLMXB until the companion has expanded off the main
sequence, if at all. These systems thus cannot produce such binaries
with unevolved companions. If on the other hand the binary is
relatively narrow, so that the primary's envelope engulfs the
companion as it expands, we must ask if the latter can survive this
common--envelope (CE) evolution. Friction between the star and the
envelope will remove orbital energy and help to unbind the
envelope. If the black--hole progenitor, its compact He core and its
envelope have masses $M_{1p}, M_c, M_e$, and $R_{1p}$ is its radius,
then equating the loss of orbital energy to the binding energy of the
envelope in the standard way shows that
\begin{equation}
{GM_{1p}M_e\over \lambda} = \alpha\biggl({GM_cM_{2i}\over 2a_f} -
{GM_{1p}M_2\over 2a_0}\biggr)
\label{ce}
\end{equation}
where $M_{2i}$ is the companion mass at this point, $a_0, a_f$ are the
initial and final orbital separations, and $\lambda, \alpha$ are the
usual weighting and efficiency parameters. For a low--mass companion
we have $M_{2i} << M_{1p}, M_c, M_e$, and the final separation obeys
\begin{equation}
a_f < {\alpha\lambda\over 2}{M_c\over M_1}{M_{2i}\over M_e}R_1 << R_1.
\label{ce2}
\end{equation}
In general the final separation is too small for the companion star to
fit inside it, so the stars merge rather than forming a binary. The
small mass ratio meant that there was insufficient orbital energy
available to eject the black--hole progenitor's envelope.  Black--hole
X--ray binaries must have initial mass ratios $q_i = M_{2i}/M_1$ above
some limiting value, and thus post--CE secondary star masses $M_{2i}$
above some minimum value (
The uncertainties inherent in any discussion
of CE evolution do not allow a precise answer, but it is plausible
that this effect limits $q_i$ to values $\ga 0.1$ and thus $M_{2i}$ to
values $\ga 1\msun$. This creates a real possibility that
this star has time to undergo significant nuclear evolution before
getting into contact and starting to transfer mass (see
Fig. \ref{taunucfig}). The first paper to consider this type of
evolution was by Pylyser \& Savonije (1988).

After the CE phase, the fate of the binary is determined by a
competition between various processes which all tend to bring the
secondary into contact. The first is nuclear evolution: the star is
changing its core composition and eventually increasing its radius on
a nuclear timescale $t_{\rm nuc}$. The other timescales $t_{\rm MB},
t_{\rm GR}$ describe orbital angular momentum loss via magnetic
braking and gravitational radiation respectively, which shrink the
binary separation. If $t_{\rm nuc}$ is always shorter than the other
timescales, the companion reaches contact as it expands away from the
main sequence, and the binary evolves towards longer periods $P \ga 1
- 2$~d with mass transfer driven by nuclear expansion. This is the
origin of the long--period LMXBs we shall discuss in the next
Section. If instead the timescale for orbital angular momentum loss
(in practice $t_{\rm MB}$) is shorter than the nuclear timescale
$t_{\rm nuc}$, the binary shrinks and reaches contact at a period of a
few hours. It becomes a short--period LMXB, with mass transfer driven
by angular momentum loss as discussed earlier in this Section. However
$t_{\rm nuc}$ is only slightly longer than $t_{\rm MB}$ for the masses
$M_{2i} \ga 1\msun$ we have inferred for black--hole binaries. (Note
that magnetic braking is probably ineffective [i.e. $t_{\rm MB}
\rightarrow \infty$] for masses $M_{2i} \ga 1.5\msun$.)  If the
post--CE separation is narrow (i.e. just wide enough to avoid a
merger) the binary will reach contact with the secondary still on the
main sequence, and produce a persistent BH + MS system (Fryer \&
Kalogera, 2001). Such special initial conditions must make these
systems intrinsically rare. In addition they will be very hard to
identify, as it is difficult to measure a dynamical mass in a
persistent system; they could lurk undetected among LMXBs which do not
show Type I X--ray bursts. For the majority of systems, the shrinkage
towards contact after the CE phase takes a noticeable fraction of
$t_{\rm nuc}$, and the companion is likely to be significantly
nuclear--evolved when it comes into contact. At this point further
nuclear evolution is frozen, as the star is losing mass on a timescale
($t_{\rm MB}$ or $t_{\rm GR}$) which is shorter than $t_{\rm
nuc}$. This argument suggests that we should expect most short--period
black--hole LMXBs to have chemically evolved secondaries, accounting
for the high fraction of transients among them.

For short--period neutron star LMXBs the question is obviously more
delicate: a majority of them appear to be persistent, and thus could
have completely unevolved companions, but there is a non--negligible
fraction of transients. It is now well understood (Kalogera \&
Webbink, 1996) that the formation of a neutron--star LMXB is an
extremely rare event, requiring highly constrained initial
parameters. This results chiefly from the requirement to keep the
binary intact when the neutron--star progenitor explodes. King \& Kolb
(1997) nevertheless found that a significant fraction of short--period
neutron--star binaries would have evolved companions and thus be
transient if the supernova explosion was assumed fairly symmetrical,
and Kalogera et al. (1998) extended this result to the case of
significantly anisotropic supernovae in which the neutron star
receives a kick.

\subsection{Post--minimum SXTs}

We have seen that a large fraction of neutron--star LMXBs apparently
have fairly unevolved companions, and are persistent systems in the
$\sim 2 - 10$~hr period range as expected from
Fig. (\ref{ulifig}). However this Figure also shows that such systems
are likely to become transient after passing the predicted minimum
period $\sim 80$~min for this type of `CV--like' binary evolution. In
the CV case, post--minimum systems are generally regarded as
undetectable owing to their faintness. However straightforward
application of the ideas of the last two Sections shows that
post--minimum NSLMXBs could very well be detectable (King, 2000). They
are likely to have outbursts reaching $\sim 10^{37}$~erg~s$^{-1}$
before declining on an e--folding timescale of a few days. In effect,
these systems bring themselves to our notice by saving up their rather
feeble mass transfer rates $\sim 10^{-11}\msun$~yr$^{-1}$ and using
them to produce X--rays for only about 1\% of their lifetimes, but
with of course 100 times the luminosity. The population of faint
transients found by {\it Beppo}SAX in a $40^{\circ}\times 40^{\circ}$
field around the Galactic centre may be drawn from this group. Almost
all of these are known to contain neutron stars, signalled by Type I
X--ray bursts. Further systematic study of this population may have
much to tell us about binary evolution in the Galaxy.

\subsection{On/off transients}

Some short--period neutron--star SXTs show variability unlike any
other LMXBs. In particular at least two short--period transients, EXO
0748--68 and GS 1826--24 seem essentially to have simply `turned on',
i.e. they were undetected for the first $\sim 30$~yr of X--ray
astronomy, but have remained `on' ever since their discovery. In
compensation, several short--period transients, e.g. X2129+470,
X1658-298, have also been observed to turn off during the same
time. These on/off transitions cannot result from disc instabilities,
as the mass involved is too great. Taking the bolometric luminosity
of X0748-678 as $\sim 10^{37}$~erg~s$^{-1}$, a 10\% efficiency of
rest--mass energy conversion requires its neutron star to have
accreted $\ga 5\times 10^{25}$~g in since 1985 when it was observed to
turn on. However the maximum surface density allowing a disc to remain in
the quiescent state leads to a maximum possible quiescent disc mass
\begin{equation}
M_{\rm disc,\ max} \simeq 10^{-8}R^3\ {\rm g}
\label{discmass}
\end{equation}
(e.g. King \& Ritter, 1998) where $R$ is the outer disc radius in cm.
The 3.82~hr period implies a total binary separation of only $9\times
10^{10}$~cm, and thus $R \la 5\times 10^{10}$~cm (assuming a disc
filling 90\% of the Roche lobe, with a conservative mass ratio
$M_2/M_1 \ga 0.1$).  From (\ref{discmass}) this gives $M_{\rm disc,\
max} \la 1.3\times 10^{24}$~g, far smaller than the mass accreted
since 1985. This can only have come from the companion star, implying
stable disc accretion during the `on' state.

The fact that these systems are all known (from the presence of Type I
X--ray bursts) to contain neutron stars offers a suggestive answer. We
see from Fig. (\ref{nslmxbfig}) that even neutron--star systems with
quite evolved secondaries only just contrive to lower their mass
transfer rates sufficiently to become transients, typically in the
period range 5 -- 10~hr. Evidently some of these systems make
transitions across this stability curve for intervals which are
shortlived in evolutionary terms, but long enough to account for the
observed turn--on as a persistent LMXB. This is quite similar to the
behaviour of a group of CVs known as Z~Cam stars, where dwarf nova
behaviour is from time to time suspended as the systems enters a
`standstill' (see e.g. Warner, 1995). This behaviour can plausibly be
ascribed (e.g. King \& Cannizzo, 1998) to starspot activity on the
secondary star; during outbursting epochs, starspots block enough of
the mass transfer region near the inner Lagrange point that the mass
transfer rate is reduced to a value slightly below the critical one
for disc instability (for the unirradiated discs in CVs). During
standstills, enough of these starspots disappear that the mass
transfer rate now reaches the regime for stable disc accretion.

The secondaries in short--period LMXBs are probably magnetically
active like those in CVs, so it is plausible that a similar effect
could cause LMXBs to move between transient and persistent
behaviour. However, since such a standstill can only occur after an
outburst has triggered the transition to the hot disc state, they must
be separated by the usual very long quiescent intervals. If the
standstills are themselves also very prolonged, and the SXT outbursts
during the low mass transfer rate phase are rare or faint, the
long--term X--ray behaviour of these systems will consist essentially
of `off' and `on' states, with only short transitions (`outbursts' and
`decays') between them. This on--off behaviour is of course just what
observed for several of these systems. Accordingly we identify the
on--off transients of Table 1 as the LMXB analogues of the Z~Cam
systems.

\subsection{Long--period SXTs}
\label{lp}

\begin{figure}
  \begin{center}
    \epsfig{file=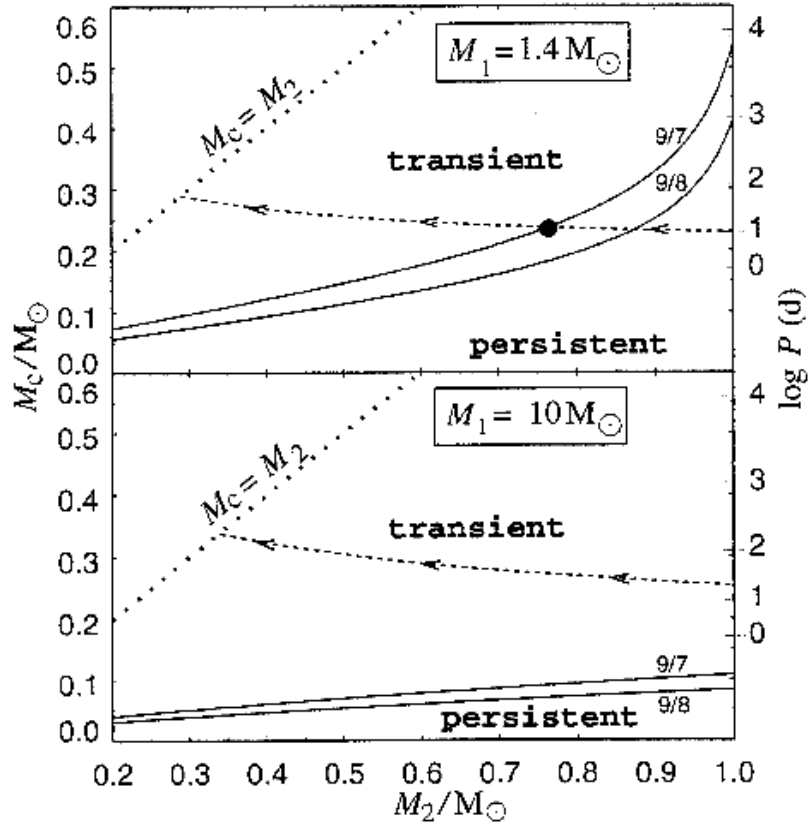, width=12cm}
  \end{center}
\caption{({\it Top}) Critical core mass versus total secondary mass
for an LMXB with a $1.4\msun$ neutron star primary. The two solid
curves give the boundaries between persistent and transient behaviour
for disc scale heights varying as $H \propto R^{9/7}$ and $R^{9/8}$
respectively. The scale on the right--hand axis gives the approximate
binary period at which the companion would fill its Roche lobe. Also
shown ({\it dashed curve}) is the track in the $M_c - M_2$ plane
followed by a binary driven by nuclear expansion of the
companion. This star has core and total masses $M_c(0) = 0.23\msun$
and $M_2(0) = 1.5\msun$ at the start of mass transfer. The positions
of known systems Sco X--2, LMC X--2, Cyg X--2 and V395 Car,
corresponding to the minimum allowed secondary masses for persistent behaviour
are all very close to the filled circle. The maximum secondary mass
for the neutron star transient GRO J1744--2844 is again close to
$0.75\msun$ (for slope 9/7) or close to $0.87\msun$ (for slope
9/8). ({\it Bottom}) Same as at top, but for a $10\msun$ primary. All
these systems are transient. Figure from King et al. (1997).}
\label{evoltrans}
\end{figure}

For orbital periods $\ga 1 - 2$~d mass transfer in LMXBs must be
driven by the nuclear expansion of the secondary. This star is a
low--mass subgiant or giant whose radius and luminosity are determined
almost purely by the mass of its helium core, quite independently of
its total mass. A simple analytic prescription for this (Webbink,
Rappaport \& Savonije, 1983; King, 1988; see also Ritter, 1999) shows
that the mass transfer rate is
\begin{equation}
-\dot M_2 \simeq 4.0 \times 10^{-10}P_{\rm d}^{0.93}m_2^{1.47}\ \msun\
 {\rm yr}^{-1}
\label{mdotev}
\end{equation}
where $P_{\rm d}$ is the orbital period measured in days and $m_2$ the
total secondary mass in $\msun$. Using this together with the
stability criterion (\ref{sxt}) King et al. (1997) showed that most
such systems must be transient according to the criterion (\ref{sxt}):
see Fig. \ref{evoltrans}. The only exceptions to this statement are
neutron--star systems where the secondary has lost relatively little
of its envelope and still has a mass $M_2 \ga 0.8 - 0.9\msun$. The
reason for this propensity of long--period systems to be transient is
clear: the accretion disc is so large that central irradiation cannot
keep its outer edge ionized.

This result has a number of important consequences. First, taken
together with the results of the previous subsection, it shows that
persistent sources are if anything the exception among LMXBs, being
largely confined to a group of short--period neutron--star sources
with fairly unevolved companions (see Table \ref{transpers}).

\begin{table}
\caption{Transient and persistent behaviour among LMXBs as predicted
by the disc instability picture} 

\begin{tabular}{llll}\hline
accretor & companion                      & $P \la 12$~hr       &  $P \ga 12$~hr \\ \hline
                            &                     &            \\
neutron star & unevolved    & persistent NSLMXB,  &       ------ \\
             &              & faint SXT after      &       \\
             &              & minimum period      &         \\   
             &              &                     &          \\
neutron star & evolved      & persistent NSLMXBs                  & SXT unless \\
             &              & plus some SXTs and    & $M_2 \ga 0.8-0.9\msun$,\\
             &              & on/off transients   & progenitors of    \\
             &              & for $P \sim 5-10$~hr & wide PSR binaries \\
                            &                     &          \\
black hole & unevolved      & formation very rare;  &
                   ------ \\
           &                & persistent BHLMXB   & \\
                            &                     &          \\

black hole & evolved        & SXT                 & SXT, microquasar \\

\end{tabular}
\label{transpers}
\end{table}

Second, we see from (\ref{mdotev}) that essentially all LMXBs with
periods longer than a day or so must have mass transfer rates $\ga
10^{-10} \msun\ {\rm yr}^{-1}$. In transients, almost all of this mass
must attempt to accrete on to the neutron star or black hole during
outbursts. With typical duty cycles $\la 10^{-2}$ it follows that the
outburst accretion rates must be $\ga 10^{-8} \msun\ {\rm yr}^{-1}$,
and so at or above the Eddington limit for a neutron star. For longer
orbital periods or smaller duty cycles the rates are still higher, and
will in general reach those corresponding to the Eddington limit for a
black hole. These predictions agree with observations of SXT
outbursts: for example V404 Cygni ($P_{\rm d} = 6.47$) had a peak
outburst luminosity $\sim 10^{39}$~erg~s$^{-1}$ (see Tanaka \& Lewin, 1995).

Third, the endpoint of this binary evolution is detectable in many
cases, and gives us a test of the theory outlined here. As the
companion burns more hydrogen in the shell source surrounding its
helium core, the mass of the latter grows, expanding the envelope,
increasing the binary period, and driving mass transfer at an
increasing rate (cf eq \ref{mdotev}). Eventually all of the envelope
mass will be used up, some added to the helium core but most
transferred to the compact accretor. We are left with a binary
consisting of the low--mass helium white dwarf core of the companion
in a wide orbit with the `bare' accretor. The orbital period is given
directly by the helium core mass, which specifies the envelope size
just before the latter was lost, and thus the period via Roche
geometry (Savonije, 1987). If the accretor was a neutron star it may
have been spun up by accreting angular momentum along with mass. This
spinup can cause the neutron star to turn on again as a radio pulsar,
a process known as recycling (Radhakrishnan \& Srinivasan,
1982). However if almost all the mass reaches the neutron star in
super--Eddington outbursts, the efficiency of both mass and angular
momentum gain will be extremely low. This effect may prevent the
neutron star spinning up to millisecond periods in systems with a
final period $\ga 100$~d if the duty cycle is $\la 10^{-2}$ (Li \&
Wang, 1998; Ritter \& King, 2001). This may account for the otherwise
suprisingly slow spin rates of some pulsars in long--period binaries.
In addition it appears that there are no millisecond pulsars in wide
circular binaries with periods $\ga 200$~d (see e.g. Table 1 in Taam
et al., 2000). However the dearth of such binaries may also reflect
formation constraints (Willems \& Kolb, 2002).

Fourth, from the arguments above it appears that radio pulsars in wide
circular binaries must descend from transients with shorter orbital
periods. Yet although at least 10 pulsar binaries with periods longer
than 50~d are known (Taam et al., 2000), we do not know of a single
neutron--star SXT with an orbital period longer than 11.8~d
(GRO~J1744-28), even though we would have expected X--ray satellites
to see outbursts from such systems anywhere in the Galaxy within the
last 30~yr. One obvious reason for this is that the outbursts are very
rare. If this is the sole reason for our failure to see outbursts,
Ritter \& King (2001) estimate that the recurrence times of the
outbursts must be at least 300~yr, and probably considerably
longer. The idea of such long recurrence times gets strong independent
support from observations of the quiescence of the long--lasting
transient KS~1731-260 (Wijnands et al. 2001, Rutledge et
al. 2002). Here the neutron star is seen to be so cool that a
considerable time, perhaps $10^3$~yr, must elapse between outbursts.

\subsection{Transient outbursts in high--mass systems}

So far this Section has dealt with outbursts in LMXBs. In general
outbursts do not occur in high--mass X--ray binaries. The obvious
reason for this is that the companion star is itself a potent
ionization source, and is able by itself to keep the accretion disc in
the hot state. If it has effective temperature $T_*$ and radius $R_*$
then the irradiation temperature on the surface of a disc element at
distance $R >>R_*$ from it is
\begin{equation}
\biggl({T_{\rm irr}\over T_*}\biggl)^4 \simeq 
{2\over 3\pi}\biggl({R_*\over R}\biggr)^3(1-\beta)
\label{as}
\end{equation}
where $\beta$ is the albedo (e.g. Frank et al., 2002, eq. 5.103; note
that the star is an extended source of irradiation since its radius
$R_*$ is much larger than the local scaleheight $H$ of the disc). For
a disc around a compact star orbiting the massive star in a binary
with a circular orbit this gives
\begin{equation}
T_{\rm irr} = 6900T_{30}R_{10}^{3/4}M_{10}^{-1/4}P_{10}^{-1/2}\ {\rm K}
\label{trad}
\end{equation}
where $T_{30}, R_{10}, M_{10}, P_{10}$ are $T_*, R_*$ and the binary
total mass $M$ and period $P$ in units of $3\times 10^4~{\rm K},
10\rsun, 10\msun$ and 10~d respectively, and we have taken
$(1-\beta)^{1/4} \simeq 1$. 

Comparing $T_{\rm irr}$ with $T_{\rm H} \simeq 6500$~K, this equation
suggests that outbursts will be suppressed in HMXBs with O or early B
primaries ($T_* \ga 30,000$~K, $R_* \sim 20 - 30\rsun$) unless the
binary period is longer than $\sim 10$~d. This agrees with the fact
that outbursts are not seen in most supergiant X--ray binaries with
known orbital periods.

However outbursts may be possible for systems with longer orbital
periods or high eccentricities $e$, since at apastron a factor
$(1+e)^{-3/4}$ appears on the rhs of eq. (\ref{trad}). Both
possibilities occur in Be X--ray binaries. Here the accretion disc is
replenished by a burst of mass transfer as the accretor (apparently
always a neutron star) passes close to the Be star's equatorial
disc. This burst, or the change in the gravitational potential felt by
the disc, is probably the cause of the outbursts usually observed near
periastron. Evidently there might in some cases be a second outburst
near apastron as the disc is allowed to cool. However the accretion
disc here is clearly not in a steady state, so numerical simulations
will be needed to check this idea.

\section{Quiescent transients and black hole horizons}
\label{hor}

One of the main motivations for studying accretion flows is to learn
more about the accreting objects. We have seen in the earlier Sections
that there are very strong indications that many compact binaries,
particularly SXTs, do contain black holes. However all of this
evidence is indirect: black holes are a consistent solution, rather
than a required one. It would be very interesting to discover {\it
direct} evidence for the defining property of a black hole, namely the
lack of a stellar surface. If the systematic difference ($n = 2$ or
$1$) in the irradiation law (eq. \ref{irr}) for black--hole and
neutron--star SXTs had held up this would have provided such
evidence. However we saw in Subsection \ref{sp} that there is little
reason to believe this.

A quite separate argument for a systematic BH/NS difference (Narayan
et al., 1997; Garcia et al., 2001) uses the idea (see Subsection
\ref{ad}) that an ADAF on to a black hole will be inherently fainter
than the same flow on to a neutron star, because the advected energy
is released at the stellar surface in the latter case.  As we have
seen, dynamical mass determinations suggest that some SXTs contain
black holes, while others contain neutron stars. There is some
observational evidence that the former systems are systematically
fainter than the latter in quiescence, as expected if indeed the two
groups have similar ADAFs in this phase.

However the last requirement is very strong, even granted that ADAFs
actually occur in quiescence, which is not entirely settled. We have
seen in Section \ref{occob} that there are systematic differences
between black--hole and neutron--star SXTs evolution. Even at similar
orbital periods, the two groups probably have different mean mass
transfer rates. This in turn may lead to differing outburst/quiescent
behaviour, undermining the assumption of similar ADAFs in the two
cases.  It is clear that much more work is needed on these effects if
quiescent transients are to provide direct evidence for black hole
horizons.

\section{Ultraluminous X--ray sources}

It has been known for more than 20 years that some external galaxies
contain X--ray sources outside their nuclei whose luminosities exceed
the Eddington limit for a $1\msun$ object (Fabbiano, 1989). These
ultraluminous X--ray sources (ULXs) have attracted considerable
interest in recent years (see e.g. Makishima et al., 2000 and
references therein) partly because one simple way of evading the
Eddington limit constraint is to assume larger black hole masses than
are generally found as the endpoints of stellar evolution
(e.g. Colbert \& Mushotzky, 1999; Ebisuzaki et al., 2001; Miller \&
Hamilton, 2002). Such intermediate--mass ($ \sim 10^2 - 10^4\msun$)
black holes are an ingredient of some pictures of galaxy formation
(e.g. Madau \& Rees, 2001), and thus raised the hope that ULXs might
represent such a population.

However it now appears that although individual ULXs might conceivably
harbour intermediate--mass black holes, this cannot be true of the
class as a whole. Instead ULXs are probably in the main X--ray
binaries in rather extreme evolutionary phases. They offer exciting
insight into many of the topics discussed in this book.

\subsection{The nature of the ULX class}

There are several lines of argument suggesting that the ULX class
involves stellar--mass accretors. These are both negative and
positive. The negative arguments concern the difficulties of forming
and then feeding intermediate--mass objects. King et al. (2001)
summarize several of these. A black hole of $10^2 - 10^4\msun$ cannot
result from current stellar evolution, as stars of $\ga 100\msun$ are
subject to huge mass loss if they have any significant metal content,
and rapidly reduce their masses to quite modest values before
producing black holes. Primordial stellar evolution (i.e. with
hydrogen and helium alone) can produce black holes with such masses,
but then the question of feeding the hole with accretion becomes
critical. As we have seen in Subsection (\ref{sp}), black--hole
binaries must either be born with separations so wide that reaching
contact at all is problematical, or have initial mass ratios above a
minimum value $q_i \ga 0.1$. With black--hole masses $M_1 > 100\msun$
the companion must have $M_{2i} \ga 10\msun$ and thus long ago have
become a compact object itself. Various other routes to making
intermediate--mass black holes have been suggested, often invoking
mergers within globular clusters. Again the process is rather
delicate, as the merged object must not attain the rather low space
velocity required to escape the cluster before its mass has built up
to the required value. ULXs are not observed to be members of globular
clusters, and indeed their incidence is often associated with recent
star formation, so the hole must eventually be ejected from the
cluster. There is then again the problem of finding a companion to
supply the hole with mass: the hole apparently did not achieve this
feat in the cluster, even given the high stellar density, but must
nevertheless manage it in the field. At the very least these
difficulties suggest that the efficiency of finding a companion and
thus turning the system on as a ULX must be rather low. The observed
numbers of ULXs found in star--forming systems such as the Antennae
($\sim 10$) therefore demand rather high formation rates for
intermediate--mass black holes if they are to explain the ULX class as
a whole.

In addition to these negative arguments, there are some positive
arguments favouring a stellar--mass black hole origin for ULXs.
First, in most cases their X--ray spectra are consistent with thermal
components at $kT \sim 1- 2$~keV. This is a natural temperature for a
stellar--mass object (see Fig. \ref{bhbb}). In addition X-ray spectral
transitions typical of such sources are observed in ULXs (e.g. Kubota
et al., 2001). Second, many ULXs, though not all, are close to regions
of star formation (Zezas, Georgantopoulos \& Ward 1999; Roberts and
Warwick 2000, Fabbiano, Zezas \& Murray, 2001; Roberts et al.,
2002). This is consistent with ULXs being the extreme end of an HMXB
population formed in such regions. Third, optical identifications
(e.g. Goad et al., 2002) are consistent with HMXBs.

On the basis of these arguments King et al. (2001) suggested that most
ULXs were probably mildly (factors $\la 10$) anisotropically emitting
X--ray binary systems accreting at close to the Eddington value. This
allows apparent luminosities up to $\sim 10^{40}$~erg~s$^{-1}$,
compatible with the great majority of claimed ULXs. At first sight
this seems to require large numbers of unseen sources, and thus a high
birthrate. However this is not so, as the following analysis shows.

We assume that a compact object of mass $M_1$ accretes from a mass
reservoir (e.g. a companion star) of mass $M_2$.  We denote the mean
observed number of ULXs per galaxy as $n$, the beaming factor as $b$
($ = \Omega/4\pi$, where $\Omega$ is the solid angle of emission), the
duty cycle (= time that the source is active as a fraction of its
lifetime) as $d$, and define an `acceptance rate' $a$ as the ratio of
mass accreted by $M_1$ to that lost by $M_2$, i.e. the mean accretion
rate $\dot M_1 = a(-\dot M_2)$. We further define $L_{\rm sph}$ as the
apparent X--ray (assumed bolometric) luminosity of a source, given by
the assumption of isotropic emission, and let $L_{40} = L_{\rm
sph}/10^{40}\ {\rm erg\ s}^{-1}$. From these definitions it follows
that the luminosity
\begin{equation}
L = bL_{\rm sph} = 10^{40}bL_{40}\ {\rm erg\ s}^{-1}
\label{l}
\end{equation}
and the minimum accretor mass if the source is not to exceed the Eddington
limit is
\begin{equation}
M_1 \ga 10^2bL_{40}\msun.
\label{m}
\end{equation}
The total number of such sources per galaxy is
\begin{equation}
N = {n\over bd}
\label{n}
\end{equation}
with a minimum mean accretion rate during active phases of
\begin{equation}
\dot M_{\rm active} = {\dot M_1\over d} = -{\dot M_2a\over d} >
10^{-6}bL_{40}\ \msun\ {\rm yr}^{-1}. 
\label{mdoto}
\end{equation}
The mass loss rate from $M_2$ is thus
\begin{equation}
-\dot M_2 >
10^{-6}{bd\over a}L_{40}\ \msun\ {\rm yr}^{-1},
\label{mdot}
\end{equation}
and the lifetime of a source is
\begin{equation}
\tau = -{M_2\over\dot M_2} \la
10^{6}{m_2a\over bdL_{40}}\ {\rm yr},
\label{tau}
\end{equation}
with $m_2 = M_2/\msun$, leading to a required birthrate per galaxy 
\begin{equation}
B = {N\over\tau} \ga {n\over bd}.{bdL_{40}\over 10^6m_2a} = 
10^{-6}{nL_{40}\over m_2a}\ {\rm yr}^{-1}.
\label{b}
\end{equation}
The important point to note here is that the required birthrate is
independent of beaming (and duty cycle): the greater intrinsic source
population $N$ required by $bd < 1$ (cf eq. \ref{n}) is compensated by
their longer lifetimes (cf eq. \ref{tau})

A possible alternative to the idea of mild anisotropy as an
explanation for ULXs was proposed by Begelman (2002), who suggested
that a magnetized accretion disc might allow luminosities which were
genuinely super--Eddington by factors up to $\sim 10$ (see also
Shaviv, 1998, 2000). An observationally--motivated objection to this
is the existence of neutron stars which have apparently passed through
phases of super--Eddington mass transfer without showing signs of
significant mass or angular momentum gain, as we might expect if
super--Eddington accretion were allowed. The difficulty in spinning up
neutron stars in wide circular binaries (Subsection \ref{lp} above) is
an example.

Strong confirmation of the idea the ULXs represent a population of
stellar--mass X--ray binaries comes from work by Grimm et
al. (2002). They show that the cumulative luminosity functions of
nearby starburst galaxies, as well as the Milky Way and Magellanic
Clouds, can be fitted by a single form normalized by the star
formation rate (SFR), as measured by various conventional
indicators. The form
\begin{equation}
N(>L) = 5.4\times SFR\times (L_{38}^{-0.61} - 210^{-0.61})
\label{lf}
\end{equation}
is used in Figures (\ref{lffig1}, \ref{lffig2}) below, where $L_{38}$
is the X--ray luminosity in units of $10^{38}$~erg~s$^{-1}$.
\begin{figure}
  \begin{center}
    \epsfig{file=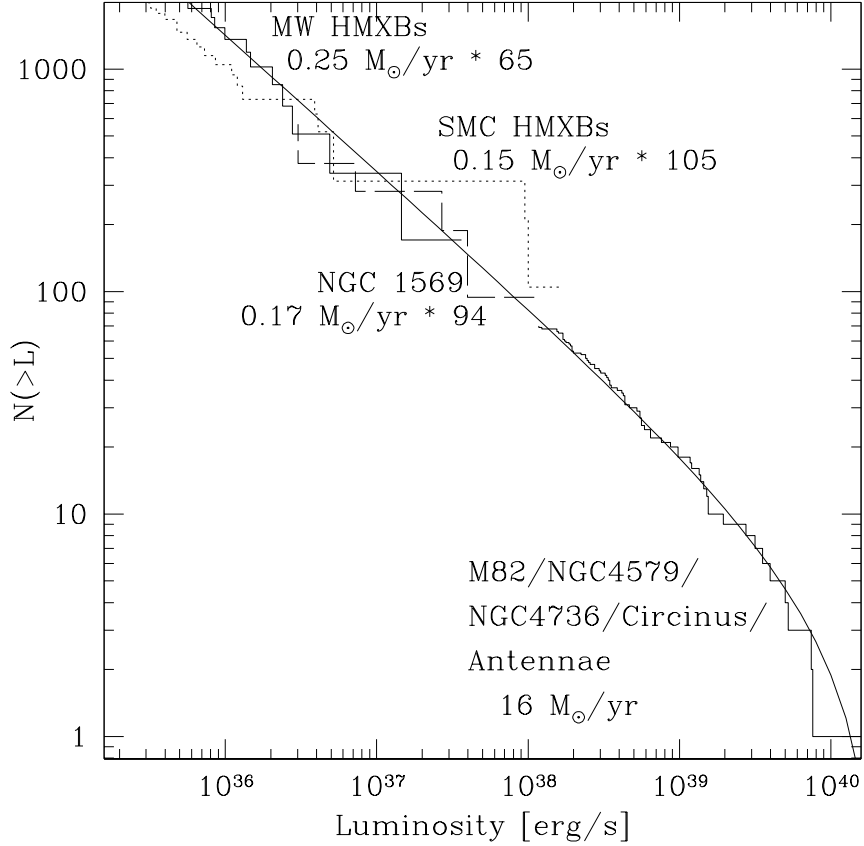, width=12cm}
  \end{center}
\caption{Combined luminosity function of compact X--ray sources in the
starburst galaxies M82, NGC 4038/9, NGC 4579, NGC 4736 and Circinus,
with a total SFR of $16\msun\ {\rm yr}^{-1}$ (above $2\times
10^{38}$~erg~s$^{-1}$0, and the luminosity functions of NGC 1569,
HMXBs in the Milky Way, and in the Small Magellanic Cloud (below
$2\times 10^{38}$~erg~s$^{-1}$). The thin solid curve is the best fit
to the combined luminosity function of the starburst galaxies only,
given by eq. \ref{lf}. Figure from Grimm et al., 2002.}
\label{lffig1}
\end{figure}
\begin{figure}
  \begin{center}
    \epsfig{file=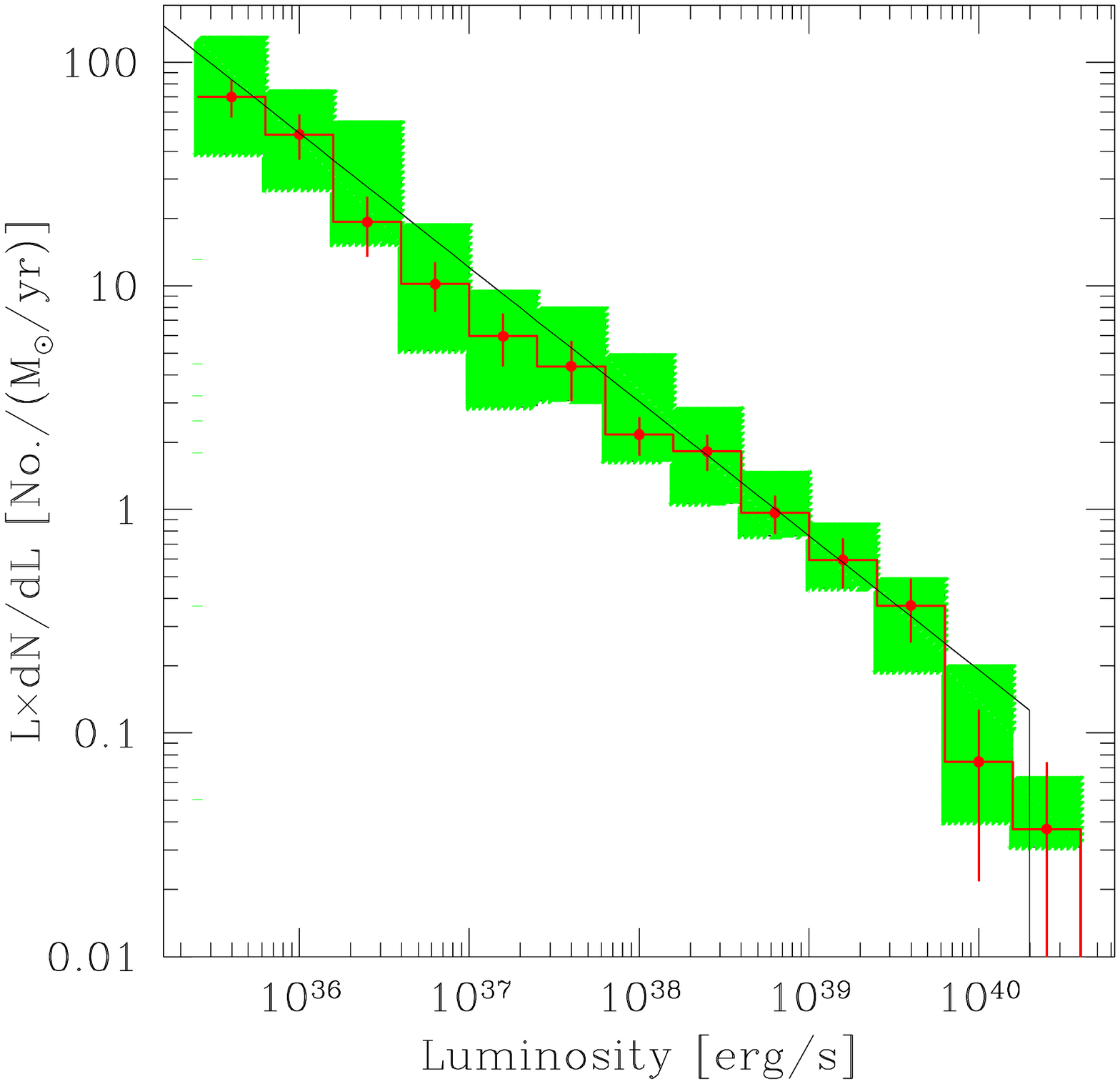, width=12cm}
  \end{center}
\caption{ Differential luminosity function obtained from the whole
sample of Grimm et al., 2002. The straight line is the same as in
Fig. \ref{lffig1}.}
\label{lffig2}
\end{figure}

These results mean that the ULX population must be some kind of
extension of the HMXB/LMXB populations contributing to the luminosity
function at lower luminosity. Note that this result is asserted only
for the ULX {\it as a class}. King et al (2001) point out that all the
arguments above still allow the possibility that {\it individual} ULXs
could involve intermediate--mass black holes. However it is probably
fair to say that at the time of writing no convincing example is
known.

\subsection{Models for ULXs}

If ULXs are X--ray binaries, we should ask what causes their unusual
appearance, and in particular their defining feature, the apparent
super--Eddington luminosity. The immediate cause appears to be a highly
super--Eddington mass inflow rate near the accretor, leading to three
characteristic features: (i) the total accretion luminosity is of
order $L_{\rm Edd}$, (ii) this is confined to a solid angle $4\pi b
\la 4\pi$, making the source apparently super--Eddington when viewed
from within this solid angle (even if it is not genuinely
super--Eddington), and (iii) the bulk of the super--Eddington mass
inflow is either accreted at low radiative efficiency, or more
probably, ejected in the form of a dense outflow, probably including
relativistic jets.

A suggested accretion flow with these features (Paczy\'nski \& Wiita,
1980; Jaroszynski et al., 1980; Abramowicz et al., 1980) postulates an
accretion disc whose inner regions are geometrically thick, and a
central pair of scattering funnels through which the accretion
radiation emerges. Note that this form of `beaming' does not involve
relativistic effects, although Doppler boosting in a relativistic jet
has also been suggested as a way of explaining the high luminosities
(Koerding et al., 2001, Markoff et al., 2001). The thick--disc plus
funnels anisotropy mechanism explicitly requires a high mass inflow
rate near the black hole or neutron star accretor, much of which must
be ejected, probably some of it in the form of a jet. (In fact the
motivation of the original papers (Jaroszynski et al., 1980,
Abramowicz et al., 1980) was to produce a geometry favouring jet
production.)

The identification with super--Eddington mass inflow rates made above
allows us to identify the likely ULX parent systems. There are two
situations in which X--ray binaries naturally have such rates: phases
of thermal--time mass transfer, and bright SXT outbursts. 
The first of these is considered extensively by King et
al. (2001) and its main features can be summarized briefly here.

Thermal--timescale mass transfer occurs in any Roche--lobe--filling
binary where the ratio $q$ of donor mass to accretor mass exceeds a
critical value $q_{\rm crit} \sim 1$. Thus all high--mass X--ray
binaries will enter this phase once the companion fills its Roche
lobe, either by evolutionary expansion, or by orbital shrinkage via
angular momentum loss. Depending on the mass and structure of the
donor, extremely high mass transfer rates $\dot M_{\rm tr} \sim
10^{-7} - 10^{-3}\msun {\rm yr}^{-1}$ ensue. SS433 is an example of a
system currently in a thermal--timescale mass transfer phase (King et
al., 2000) which has descended by this route. The idea that SS433
itself might be a ULX viewed `from the side' provides a natural
explanation of its otherwise puzzlingly feeble X--ray emission
($L_{\rm x} \sim 10^{36}$~erg~s$^{-1}$, Watson et al., 1986).

The binary probably survives the thermal--timescale phase without
entering common--envelope (CE) evolution provided that the donor's
envelope is largely radiative (King \& Begelman, 1999). Observational
proof of this is provided by Cygnus X--2 (King \& Ritter, 1999;
Podsiadlowski \& Rappaport, 2000), whose progenitor must have been an
intermediate--mass binary (companion mass $\sim 3\msun$, neutron star
mass $\sim 1.4\msun$). CE evolution would instead have engulfed the
binary and extinguished it as a high--energy source. The binary would
probably have merged, producing a Thorne--$\dot{\rm Z}$ytkow object.

The birthrates of intermediate and high--mass X--ray binaries are
compatible with the observed numbers of ULXs: King et al. (2001) show
that the birthrates required to explain the latter are independent of
the dimensionless beaming and duty--cycle factors $b, d$. For massive
systems the thermal timescale lasts longer than the preceding
wind--fed X--ray binary phase; the fact that there are far fewer
observed ULXs than massive X--ray binaries must mean that the beaming
and duty--cycle factors obey $bd << 1$. This picture also explains the
observed association of ULXs with star formation. There is in addition
some evidence that the ULXs in the Antennae are on average slightly
displaced from star clusters, suggesting that they have acquired
significant space velocities as a result of a recent supernova
explosion, just like HMXBs (Zezas et al., 2002). If this is correct it
is a direct demonstration that the masses of ULX systems are not
unusually high.

While thermal--timescale mass transfer probably accounts for a
significant fraction of observed ULXs, bright SXT outbursts will also
produce super--Eddington accretion rates, and are the only possibility
for explaining the ULXs observed in elliptical galaxies (King, 2002).
SXT outbursts in long--period systems are an attractive candidate
because they are both bright and long--lasting.

SXT outbursts in systems with such periods are complex because of the
large reservoir of unheated mass at the edge of the disc, which can
eventually contribute to the outburst (see Section \ref{sxtnat}). Full
numerical calculations will be needed to describe this
process. However the trends with increasing $P$ are clear: the
outbursts become longer (several decades) and involve more mass, but
the quiescent intervals increase more rapidly (several $\ga 10^3$~yr)
so that the outburst duty cycle $d$ decreases (Ritter \& King,
2001). This results in inflow rates which become ever more
super--Eddington at large $P$.
Spectacular evidence of super--Eddington accretion is provided by
GRS~1915+105, which has been in effectively continuous outburst since
1992. The observed X--ray luminosity $L_{\rm x} \ga 7\times
10^{39}$~erg~s$^{-1}$ implies that at least $\sim 10^{-6}\msun$ has
been accreted over this time, requiring a large and massive accretion
disc. In line with this, it appears that the binary is wide ($P \simeq
33$~d; Greiner, Cuby \& McCaughrean, 2001). At the reported
accretor mass $M_1 = (14 \pm 4)\msun$ (Greiner et al., 2001) there is
little doubt that the current mass inflow near the black hole is
highly super--Eddington.  Evolutionary expansion of the donor will
drive a persistent mass transfer rate $-\dot M_2 \sim 10^{-9}(P/{\rm
d})\msun{\rm yr}^{-1} \sim 3 \times 10^{-8}\msun{\rm yr}^{-1}$ (King,
Kolb \& Burderi, 1996) which is already close to the Eddington rate
$\dot M_{\rm Edd} \sim 10^{-7}\ \msun{\rm yr}^{-1}$. Given an outburst
duty cycle $d << 1$, the mean inflow rate $\sim -\dot M_2/d$ is
$>>\dot M_{\rm Edd}$. Note that we definitely do not look down the jet
in GRS~1915+105, which is at about $70^{\circ}$ to the line of sight
(Mirabel \& Rodr\'iguez, 1999), so it is quite possible that the
apparent luminosity in such directions is much higher than the
observed $L_{\rm x}$. 

The observed ULX population of a given galaxy is a varying mixture of
these thermal--timescale and transient types, depending on the star
formation history of that galaxy. Thermal--timescale SS433--like
systems should predominate in galaxies with vigorous star formation,
such as the Antennae, while ULXs in elliptical galaxies must be of the
microquasar transient type, as there are no high--mass X--ray
binaries. We therefore expect ULXs in ellipticals to be
variable. However the microquasar systems most likely to be identified
as ULXs are clearly those with the brightest and longest outbursts, so
baselines of decades may be needed to see significant numbers turning
on or off. There is some evidence of such variability from the
differences between ROSAT and {\it Chandra} observations of the same
galaxies.  The fact that none of the SXTs found in the Galaxy has
turned out to be a ULX suggests that the beaming factor $b$ must be
$\la 0.1$ for this mode of accretion. This agrees with our conclusion
above that $b << 1$ for the ULXs in ellipticals.

Evidence that the two suggested classes of ULXs do resemble each other
in similarly super--Eddington accretion states comes from Revnivtsen
et al. (2002), who report RXTE observations of an episode of
apparently super--Eddington accretion in the soft X--ray transient
V4641 Sgr. Revnivtsen et al. remark on the similarity of the object's
appearance to SS433 in this phase. One might be discouraged by the
apparent suppression of X--rays in this state. However we are
presumably outside the beam of most intense X--ray emission in both
cases: neither should actually appear as a ULX. More work is needed on
whether the X--ray spectra from these objects are consistent with
X--rays leaking sideways from the assumed accretion geometry. Direct
evidence that X--ray emission in ULXs is anisotropic is perhaps
understandably meagre, but may be suggested by the comparison of
optical and X--ray data in NGC 5204 X--1 (Roberts et al., 2002), where
low--excitation optical spectra are seen from regions close to the
ULX.

Both SS433 and the microquasars are distinguished by the presence of
jets, at least at some epochs. In SS433 the jets precess with a
164--day period, presumably because of disc warping (Pringle,
1996). If looking closely down the jet is required in order to see
high luminosities one might expect to see such periods in a class of
ULXs. This effect could for example explain the $\sim 106$~d
modulation seen in the bright source in M33 (M33 X--8, $L_{\rm x} \sim
10^{39}$erg~s$^{-1}$) by Dubus et al. (1999). However a beam as narrow
as commonly inferred ($\la 1^{\circ}$) for SS433 would give an
unacceptably short duty cycle. If instead it is not necessary to look
down the jet to see a high luminosity, this would rule out Doppler
boosting as the cause of the latter, and ULXs would not be direct
analogues of BL~Lac systems.

\subsection{Black--hole blackbodies}

The tentative conclusion at the end of the last subsection suggests
another. X--ray binaries and active galactic nuclei share the same
basic model, and so far have shown a fairly good correspondence in
their modes of behaviour. If as suggested above ULXs do not correspond
to BL~Lac systems, this may mean that we are currently missing a class
of each type: there should exist an apparently super--Eddington class
of AGN, and a set of X--ray binaries with Doppler--boosted X--ray
emission. 
\begin{figure}
  \begin{center}
    \epsfig{file=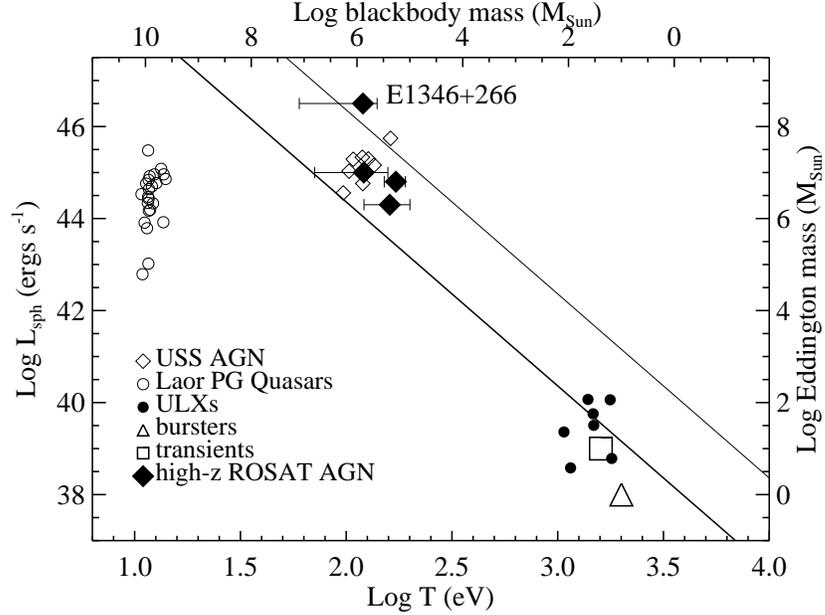, width=12cm}
  \end{center}
\caption{ Luminosity and blackbody temperature for bright X--ray
blackbody sources: ultrasoft AGN (diamonds), the Laor et al. (1997)
sample of PG quasars (open circles), ULXs (filled circles). X--ray
transients (open square) and bursters (triangle) are shown
schematically only, in the interests of clarity. The solid line is the
limit $L_{\rm sph}=L_0$. Sources below this line are compatible with
the constraints of the Eddington limit, isotropic emission and an
emission area no smaller thean the Schwarzschild radius, but must have
masses $M$ respectively above and below the values given on the rh
vertical and upper horizontal scales. Sources above the line must
either (a) violate the Eddington limit, or (b) have a significantly
anisotropic radiation pattern, or (c) emit from a region much smaller
than their Schwarzschild radii. Figure from King and Puchnarewicz
(2002) }
\label{bhbb}
\end{figure}

A tentative answer to one of these questions has recently emerged.
Many black--hole sources emit a substantial fraction of their
luminosities in blackbody--like spectral components. It is usual to
assume that these are produced in regions at least comparable in size
to the hole's Schwarzschild radius, so that a measure of the emitting
area provides an estimate of the black hole mass $M$. However there is
then no guarantee that the source luminosity (if isotropic) obeys the
Eddington limit corresponding to $M$. King \& Puchnarewicz (2002) show
that the apparent blackbody luminosity $L_{\rm sph}$ and temperature
$T$ must obey the inequality 
\begin{equation}
L_{\rm sph} < L_{\rm crit} = 2.3\times 10^{44}(T/100\
{\rm eV})^{-4}~{\rm erg~s}^{-1}, 
\label{bbedd}
\end{equation}
(where $T_{100}$ is $T$ in units of 100eV) for this to hold. This
limit is shown in Fig. \ref{bhbb}. Sources violating it must either be
super--Eddington, or radiate anisotropically, or radiate from a region
much smaller than their Schwarzschild radii. Not suprisingly, some
ULXs appear above the limit. (Note that they are not required to do
this to qualify as ULXs: the defining characteristic is simply that
their `Eddington masses' [rh scale of Figure \ref{bhbb}] are $>>
10\msun$.) The large group of AGN violating the limit are the
so--called ultrasoft AGN, which may thus be the AGN analogues of the
ULXs. The second question remains: a search for for Doppler--boosted
X--ray binaries among the ULXs may be rewarding.

\subsection{Supersoft ULXs}

Very recently a number of ULXs have been observed with very low
spectral temperatures ($\sim 50-100$~eV) and consequent photospheric
sizes ($\sim 10^9$~cm) much larger than the Schwarzschild radius of a
stellar--mass object (e.g. Mukai et al., 2002). At first sight these
might at last appear as strong evidence for the long--sought
intermediate--mass black holes. However Mukai et al. (2002) have
pointed out that accretion at rates comparable to Eddington must lead
to outflow, and shown that the opacity of the resulting wind does
imply supersoft emission with a photospheric size of this order. The
M101 source studied by Mukai et al (2002) has a supersoft luminosity
of order $10^{39}$~erg~s$^{-1}$ and so does not require anisotropic
emission for a black hole mass $\ga 10\msun$. However their analysis
is easily extended to the case that an Eddington--limited source blows
out a wind confined to a double cone of total solid angle $4\pi b$
about the black hole axis. Since this wind is the path of lowest
optical depth through the accretion flow, the radiation will escape
this way also, implying anisotropic emission once again.  Mukai et al
(2002) assume a constant velocity for the outflowing material as this
is likely to achieve escape velocity and coast thereafter. This leads
to an equivalent hydrogen column from radius $R$ to infinity of $N_H =
\dot M_{\rm out}/4\pi bvR$ and thus (assuming Compton scattering
opacity) a photospheric radius
\begin{equation} 
R_{\rm ph} = {3\times 10^8\over bv_9}\dot M_{19}\ {\rm cm} 
\label{wind}
\end{equation} 
where $v_9$ is $v$ in units of $10^9$ cm~s$^{-1}$ and $\dot M_{19}$ is
the outflow rate in units of $10^{19}$~g~s$^{-1}$, the Eddington
accretion rate for a $10\msun$ black hole. Clearly we can again
interpret such supersoft ULXs in terms of stellar--mass black
holes. 

It is worth noting that the presence of a photosphere of this kind
seems inevitable in any source accreting significantly above the
Eddington accretion rate $\dot M_{\rm Edd}$. A completely general
calculation (Pounds et al., 2003) shows that
\begin{equation}
{R_{\rm ph}\over R_{\rm s}} = {1\over 2\eta b}{c\over v}{\dot M_{\rm
out}\over \dot M_{\rm Edd}}\simeq {5\over b}{c\over v}{\dot M_{\rm
out}\over \dot M_{\rm Edd}}
\end{equation}
where we have taken the accretion efficiency $\eta \simeq 0.1$ at the
last step. Since $b \leq 1, v/c < 1$ we see that $R_{\rm ph} > R_{\rm
s}$ for any outflow rate $\dot M_{\rm out}$ of order $\dot M_{\rm Edd}$. In
other words, any black hole source accreting at above the Eddington rate
is likely to have a scattering photosphere at several $R_{\rm s}$.

\section{Conclusions}

Our picture of accretion in compact binary systems has advanced
considerably over recent years. In the past it was common to think of
these sources as relatively steady systems which accreted most of the
mass transferred to them, often from main--sequence companions, and
radiated roughly isotropically. It now seems that none of these
implicit assumptions is really justified.  Transient behaviour is
extremely widespread, to the point that persistent sources are rather
exceptional. Much of the transferred mass is not accreted at all, but
blown away from the accretor: jets are only the most spectacular
manifestation of a very widespread trait. Even short--period systems
often have significantly evolved companions, and there is little
evidence for a period gap for short--period LMXBs. Disc warping
and other effects can apparently cause many sources to radiate with
significant anisotropy. 

Despite these complicating effects there are reasons for optimism.
One can now give a complete characterization of the observed incidence
of transient and persistent sources in terms of the disc instability
model and formation constraints. X--ray populations in external
galaxies, particularly the ultraluminous sources, are revealing
important new insights into accretion processes and compact binary
evolution.

\bigskip
{\bf Acknowledgments}

I thank Ed Colbert, Juhan Frank, Hans--Jochen Grimm, Vicky Kalogera,
Uli Kolb, James Murray, Tim Roberts, and particularly Klaus Schenker
and Dan Rolfe for much help in the writing of this review. Theoretical
astrophysics research at Leicester is supported by a PPARC rolling
grant. I gratefully acknowledge a Royal Society Wolfson Research Merit
Award.

\end{document}